\documentclass[twoside,a4paper]{article}
\usepackage{a4wide,amsmath,bm,booktabs,array,parskip}
\usepackage{natbib}
\usepackage[english]{babel}
\usepackage{blindtext}
\usepackage{multirow}
\usepackage{verbatim}
\usepackage{appendix}
\usepackage{hyperref} 
\usepackage{graphicx}
\usepackage{float}
\usepackage{footmisc}
\usepackage{bbm}
\usepackage{epsfig}
\usepackage{lipsum}
\usepackage{amsmath, amsthm, amssymb, amsfonts}
\usepackage{mathbbol}
\usepackage{latexsym}
\usepackage{epstopdf,titling,url}
\usepackage[T1]{fontenc}
\usepackage[utf8]{inputenc}
\usepackage{authblk}
\usepackage{pbox}
\usepackage{longtable, tabularx}
\usepackage{array}
\usepackage{booktabs}
\usepackage[table]{xcolor}
\usepackage{multirow}
\usepackage{a4wide,amsmath,bm,booktabs,array,parskip}
\usepackage{natbib}
\usepackage[english]{babel}
\usepackage{blindtext}
\usepackage{multirow}
\usepackage{verbatim}
\usepackage{appendix}
\usepackage{hyperref} 
\usepackage{graphicx}
\usepackage{float}
\usepackage{footmisc}
\usepackage{bbm}
\usepackage{epsfig}
\usepackage{lipsum}
\usepackage{amsmath,amsthm,amsfonts}
\usepackage{amssymb}
\usepackage{mathbbol}
\usepackage{latexsym}
\usepackage{epstopdf,titling,url,array}
\usepackage[T1]{fontenc}
\usepackage[utf8]{inputenc}
\usepackage{authblk}
\usepackage{gensymb}
\usepackage{chngcntr}
\usepackage{graphicx}

 \makeatletter
 \def\@eqnnum{{\normalsize \normalcolor (\theequation)}}
  \makeatother

\newcommand{\diag}{\operatorname{diag}}

\theoremstyle{definition}

\stepcounter{secnumdepth}
\stepcounter{tocdepth}
\title{\textsf{\bf\huge A comparison of functional summary statistics to detect anisotropy of three-dimensional point patterns}}

\author[1]{\bf Farzaneh Safavimanesh\thanks{f\_safavimanesh@sbu.ac.ir}}
\author[2]{\bf Claudia Redenbach\thanks{redenbach@mathematik.uni-kl.de}}
\affil[1]{\small Department of Mathematical Sciences, Institute for Cognitive and Brain Sciences, Shahid Beheshti University, Iran}
\affil[2]{\small Department of Mathematics, Technical University of Kaiserslautern, Germany}
 
\usepackage{fancyhdr}
\begin{document}

\thispagestyle{empty}
\date{}

\maketitle

\begin{abstract} 
The growing availability of three-dimensional point process data asks for a development of suitable analysis techniques. In this paper, we focus on two recently developed summary statistics, the conical and the cylindrical $K$-function, which may be used to detect anisotropies in 3D point patterns. We give some recommendations on choosing their arguments and investigate their ability to detect two special types of anisotropy. Finally, both functions are compared on some real data sets from neuroscience and glaciology.

\end{abstract}

\section{Introduction}\label{sec:intro} 

In some situations, the spatial correlation between the points in a point pattern is not only a function of the distances between the points, but also of the direction of the vector connecting them. Classical functional summary statistics such as Ripley's $K$-function or the nearest neighbor distance distribution function fail to detect such anisotropies. Hence, there is some interest in developing methods which allow for a detection and characterization of the degree of anisotropy in a spatial point pattern. 

In the literature, the case of two-dimensional point patterns has been at the focus of interest up to now. Various approaches for anisotropy analysis have been introduced, including spectral methods  \citep{bartlett-64,MuggRen-96,Renshaw-02}, wavelet transformations \citep{ErcoleMateu-13-1,ErcoleMateu-13-2,ErcoleMateu-14}, and an anisotropy test based on the asymptotic joint normality of the sample second-order intensity function \citep{Guanetal-06}. In addition, directional versions of functional summary statistics have been introduced in \citet{OhSt-81},  \citet{StoyanBenes-91}, \citet{Stoyan-91}, and \citet{StoyStoy-94}. Moreover, \citet{MoellerHokan-14} introduced geometric anisotropic pair correlation functions. 

At least some of these methods may be transferred to the three-dimensional case in theory. In practice, however, their application might be hampered, e.g.\  by problems in finding a suitable partition of the unit sphere. Furthermore, the visualization and verification of results is more challenging.
Recently, two directional counterparts of Ripley's $K$-function for the analysis of three-dimensional point patterns were introduced. The common idea of both approaches is to replace the ball used in the definition of the $K$-function by a structuring element which is sensitive to direction. The motivation of the work in \citet{Redenbachetal-09} was to detect anisotropy introduced by the compression of a regular point pattern. For this purpose, the mean numbers of points contained in cones centered in the typical point and pointing to different directions were investigated. The motivating data sets were point patterns of bubble centers extracted from tomographic images of polar ice cores.  In contrast, \citet{plcpp-15} studied some data from neuroscience where the points are believed to be organized in linear columns. Hence, they decided to use a cylinder instead of a cone. 

These examples illustrate how the development of methods may be triggered by the particular shape of anisotropy that should be detected. In the current paper, we want to investigate the generality of the two directional versions of the $K$-function. For this purpose, we will apply them to both real and simulated point patterns with different sources and various degrees of anisotropy.

In Section~\ref{sec:data} we introduce the data sets used throughout the paper. Section~\ref{sec:summaries} defines the conical and cylindrical $K$-functions.  Based on a non-parametric isotropy test, some simulation-based recommendations on the parametrization of these summary statistics are given in Section~\ref{sec:volume}. Finally, in Section~\ref{sec:examples}, we apply these recommendations when comparing the functions in detecting the anisotropy in real pyramidal cell and ice data sets as well as realizations of models mimicking the structure of these data.  

\section{Data sets}\label{sec:data}

In this section, we introduce the data sets used for the subsequent analyses. We start with presenting the real data studied in \citet{plcpp-15} and \citet{Redenbachetal-09} providing the motivation for the development of the two versions of the directional $K$-function. To allow for an investigation of the performance of the methods under varying degrees of anisotropy, our analysis is extended to simulated data sets. The models are chosen to reproduce the type of anisotropy present in the neuroscience data and the ice, respectively.

\subsection{Real data sets}\label{sec:realdata}

\subsubsection{Pyramidal cell point patterns}
\label{sec:pyramidalData}
The first set of data consists of four samples containing the locations of the pyramidal cells from the Brodmann area 4 of the gray matter of the human brain collected by the Center for Stochastic Geometry and Advanced Bioimaging, Denmark. According to the minicolumn hypothesis in neuroscience (see e.g.\ \citet{Mountcastle-57} and \citet{RSDRMN-15}), the point patterns are expected to be anisotropic due to the linear arrangement of the cells in a direction perpendicular to the pial surface of the brain, i.e, the $xy$-plane here. For more details on these  data sets, see \citet{RSDRMN-15}. A visualization of one sample is shown in Figure \ref{fig1:dataAll}.

\subsubsection{Ice data}
\label{sec:IceData}
\begin{figure}[t]
\centering
\includegraphics[width=3.4cm, height = 2.9cm]{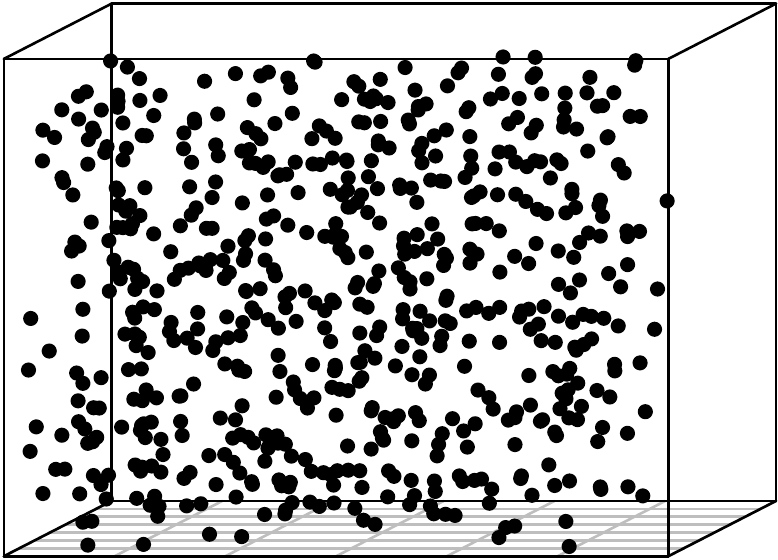}
\includegraphics[width=2.9cm]{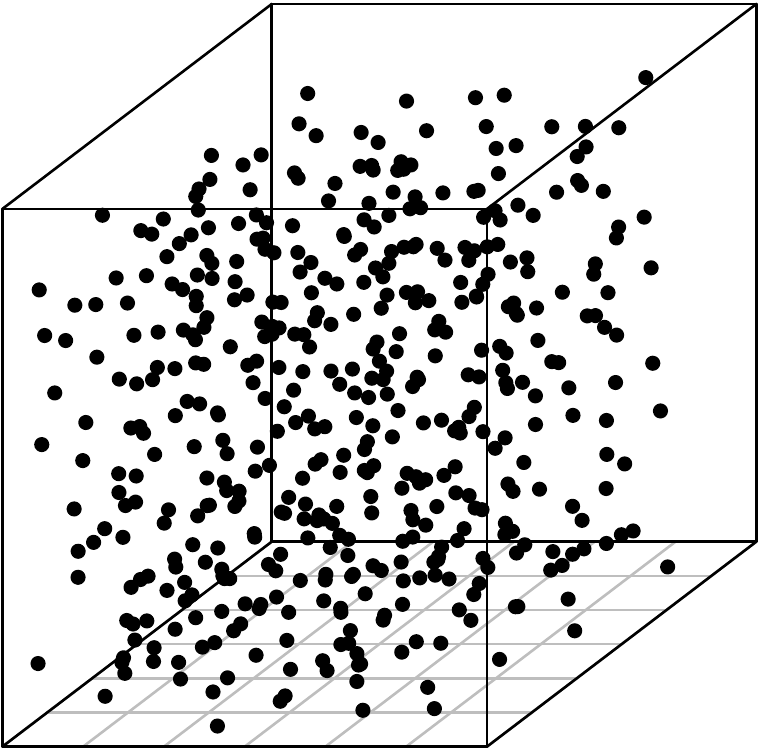}
\includegraphics[width=2.9cm]{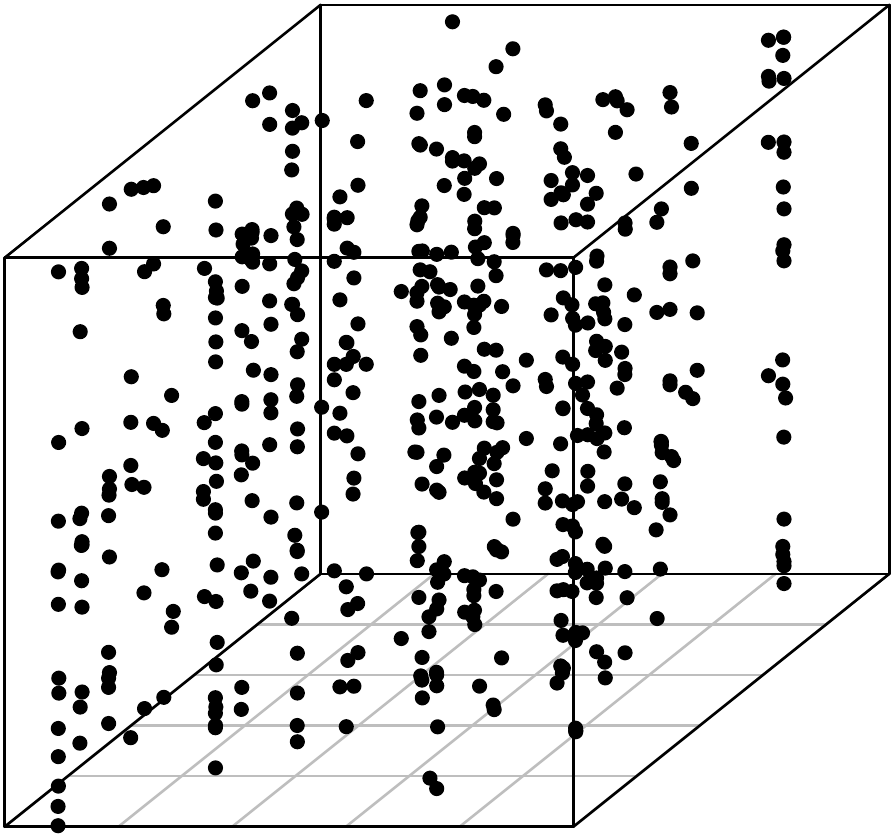}
\includegraphics[width=2.9cm]{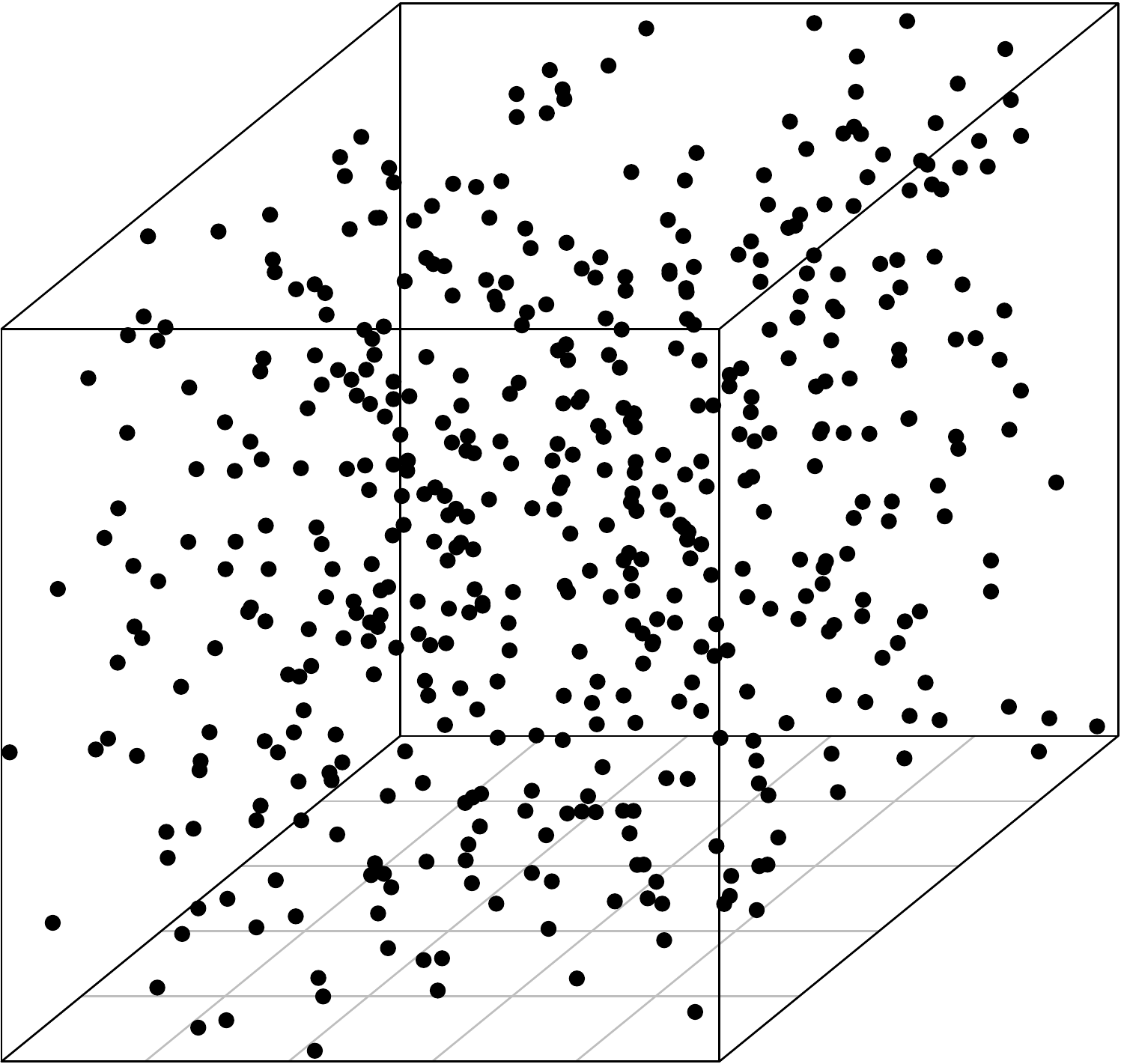}
\caption{From left to right: a sample of the pyramidal data sets within an observation window of size $508\times 140\times 320\,\mu{\text{m}}^3$, a sample of the ice data within an observation window of size $11.68\times 11.92\times 13.81$ mm$^3$, a realization of the PLCPP model for $\rho = 500, \sigma = 0.001, \rho_L = 200, \alpha = 2.5$, and the compressed center locations of random ball packing for $\rho = 500, R = 0.05, c=0.7$.
}
\label{fig1:dataAll}
\end{figure}

The second set of data consists of a subset of the samples investigated in \cite{Redenbachetal-09}. The point patterns consist of the center locations of air bubbles extracted from tomographic images of the Talos Dome ice core. The data were provided by the Alfred-Wegener-Institute for Polar and Marine Research, Bremerhaven, Germany. Details on the acquisition and the processing of the data can be found in \cite{Redenbachetal-09}. Here, we consider 14 samples taken from a depth of 505 m where the anisotropy is most prominent. The point patterns can be interpreted as realizations of a regular point process. Anisotropy is introduced by a compression of the point pattern along the $z$-axis. Due to the location of the drilling site for this ice core, isotropy within the $xy$-plane can be assumed. A visualization of one sample is shown in Figure \ref{fig1:dataAll}.

\subsection{Simulated datasets}\label{sec:simdata}

\subsubsection{Poisson line cluster point processes}
\label{sec:data-plcpp}
Motivated by the pyramidal cell data, a Cox process model called Poisson line cluster point process (PLCPP) for anisotropic spatial point processes was developed in \citet{plcpp-15}. The anisotropy of the realizations of this model is caused by linear arrangement of the points. For this purpose, we start with an anisotropic Poisson line process with intensity $\rho_L$ and a given directional distribution of lines. On each line $l_i$ contained in this process, a homogeneous Poisson process $Y_i$ with intensity $\alpha$ is independently generated. Finally, the points of the $Y_i$ are displaced in a plane orthogonal to $l_i$ by e.g.\ a zero-mean normal distribution with the standard deviation $\sigma$ yielding independent Poisson processes $X_i$ whose superposition forms the PLCPP model $X$. The parameter $\sigma$ controls the distances between the points and the lines. The intensity of $X$, i.e.\ the parameter $\rho$, is equal to the product of the intensity $\rho_L$ of the Poisson line process and the intensity $\alpha$ of the Poisson processes $Y_i$ on the lines. 

Our investigations are based on PLCPP models with intensity $\rho = 500$, $\alpha = 2.5$, and $\rho_L = 200$, where the lines are parallel to the $z$-axis. We consider a high ($\sigma = 0.001$), medium ($\sigma = 0.01$), low ($\sigma = 0.02$), and very low ($\sigma = 0.04$) degree of linearity. Figure~\ref{fig1:dataAll} shows a realization of a PLCPP model with a high degree of linearity. For the simulation study reported in the following, $m=1000$ realizations were generated for each set of parameters.

\subsubsection{Compressed regular point patterns}\label{sec:regular}

As discussed in \citet{Redenbachetal-09}, the structure of the ice data can be modelled via compression of isotropic regular point processes. To represent different degrees of regularity, we consider both a Mat\'{e}rn hard-core process (low regularity, \cite[Section 6.5.2]{Illianetal-08}) and the center locations of balls in a dense packing simulated using the force-biased algorithm (high regularity, \cite[Section 6.5.5]{Illianetal-08}). In both cases, the intensity was chosen as $\rho = 500$ and the hard core radius was $R = 0.05$. Anisotropy was then introduced by applying a volume-preserving linear transformation $T_c = \diag(1/\sqrt{c}, 1/\sqrt{c}, c), c\in [0,1]$, to these isotropic regular point patterns. This implies that the data are compressed by a factor $0<c<1$ in $z$-direction while they are isotropically stretched by a factor $1/\sqrt{c}$ in the $xy$-plane.

As in the case of the PLCPP models, $m = 1000$ realizations for each model and each set of parameters were generated within the unit cube. Different degrees of compression were realized by choosing $c=0.7,$ $0.8$, and $0.9$. Figure~\ref{fig1:dataAll} shows a realization of a point pattern obtained from a ball packing compressed by a factor $c = 0.7$. 

\section{Conical and cylindrical $K$-functions} \label{sec:summaries}

\begin{figure}[t]
\centering
\frame{\includegraphics[height=6.5cm, width=6.5cm]{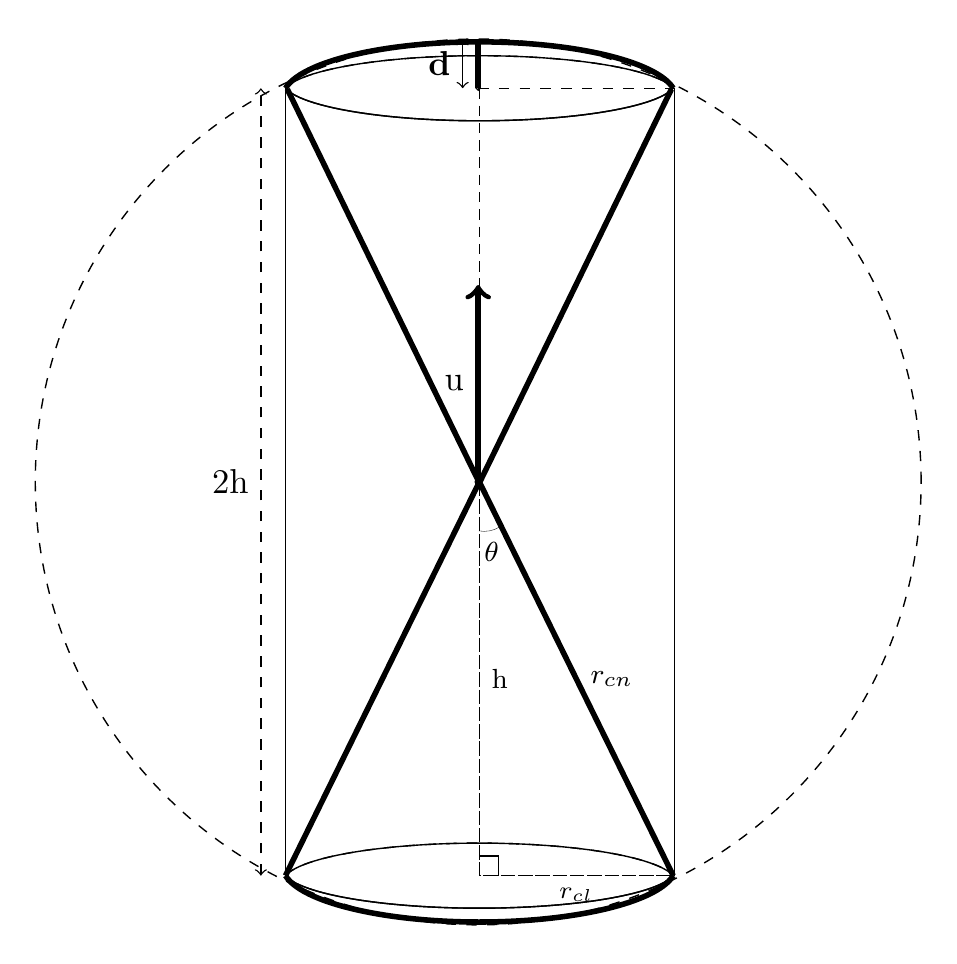}}
\caption{Structuring elements of the conical (the double cone) and the cylindrical (the cylinder) $K$-functions. 
}
\label{fig2:strucElmnt}
\end{figure}

Ripley's $K$-function is a well-known summary statistic which is defined as the mean number of further points within a circle/sphere with radius $r$ centered in a typical point of the point pattern divided by the intensity. Naturally, anisotropy cannot be detected using this function due to its symmetric structuring element. 
\citet{Redenbachetal-09} generalized the 2D directional $K$-function (see e.g.\ \citet{StoyStoy-94}) to the three-dimensional case by replacing the sector of a circle by a double cone.
 For a unit vector $\mathbf u$, the conical $K$-function is defined as 
\[ 
K_{\mathbf u, \text{cn}}(r_\text{cn}, \theta)=\frac{1}{\rho^2 |W|}\,\mathrm E\sum_{\mathbf x_1,\mathbf x_2\in
 \mathbf X}^{\not=}\mathbf1[\mathbf x_1\in W,\mathbf x_2-\mathbf x_1\in C_{\mathbf u}(r_\text{cn},\theta)],\quad 0<r_\text{cn},\,0\leq\theta\leq \frac{\pi}{2}
\]
where $\rho$ is the intensity, and $C_{\mathbf u}(r_\text{cn},\theta)$ denotes a double spherical cone in the direction $\mathbf u$ with an slant height of length $r_\text{cn}$ and an apex angle of size $2\theta$ centered in $0$ (see Figure~\ref{fig2:strucElmnt}). Briefly speaking, ${\rho}K_{\mathbf u, \text{cn}}(r_\text{cn}, \theta)$ is the mean number of further points within a cone $x_0+ C_{\mathbf u}(r_\text{cn},\theta)$ centered in a typical point $x_0$ of the point pattern. In \citet{Redenbachetal-09} the function $K_{\mathbf u, \text{cn}}$ was called directional $K$-function. Here, we will call it conical $K$-function to distinguish it from the cylindrical $K$-function introduced below.

\citet{plcpp-15} introduced a summary statistic, called the cylindrical $K$-function, to detect anisotropy of point patterns with columnar structure. It is a version of the space-time $K$-function \citep{Diggleetal-95} and is defined via
 \[ 
K_{\mathbf u, \text{cl}}(r_\text{cl},h)=\frac{1}{\rho^2|W|}\,\mathrm E\sum_{\mathbf x_1,\mathbf x_2\in
 \mathbf X}^{\not=}\mathbf1[\mathbf x_1\in W,\mathbf x_2-\mathbf x_1\in Z_{\mathbf u}(r_\text{cl},h)],\quad r_\text{cl},h>0,
\]
where $Z_{\mathbf u}(r_\text{cl},h)$ denotes a cylinder with center $0$, base radius $r_\text{cl}$, and height $2h$ in the direction $\mathbf u$ (see Figure~\ref{fig2:strucElmnt}). 
Briefly speaking, $\rho K_{\mathbf u, \text{cl}}(r_\text{cl},h)$ is the mean number of further points within a 
cylinder $x_0+Z_{\mathbf u}(r_\text{cl},h)$ centered in a typical point $x_0$ of the point pattern. For more details on the cylindrical $K$-function, see \citet{plcpp-15}. 

Ratio-unbiased non-parametric estimates of the functions are, respectively, given by 
\begin{equation}\label{e:conKest}
  \hat K_{\mathbf u, \text{cn}}(r_\text{cn}, \theta) = \frac{1}{\widehat{\rho^2}}\sum_{\mathbf x_1,\mathbf x_2\in
 \mathbf W}^{\not=} 
w(\mathbf x_1,\mathbf x_2)
  \mathbf1[\mathbf
  x_2-\mathbf x_1\in C_{\mathbf u}(r_\text{cn},\theta)].
\end{equation} 
and
\begin{equation}\label{e:conKest}
  \hat K_{\mathbf u, \text{cl}}(r_\text{cl},h) =\frac{1}{\widehat{\rho^2}}\sum_{\mathbf x_1,\mathbf x_2\in
 \mathbf W}^{\not=} 
w(\mathbf x_1,\mathbf x_2)
  \mathbf1[\mathbf
  x_2-\mathbf x_1\in Z_{\mathbf u}(r_\text{cl},h)].
\end{equation} 
where $n$ is the number of points in the point pattern, $\widehat{\rho^2}=n(n-1)/|W|^2$ is an unbiased estimate of $\rho^2$ (see e.g.\ \citet{Illianetal-08}), and $w$ is the translation edge correction factor defined as 
$$w(\mathbf x_1,\mathbf x_2) = 1/|W\cap W_{\mathbf x_2-\mathbf x_1}|$$ 
in which $W_{\mathbf x}$ denotes the translation of the observation window $W$ by the vector $\mathbf x$ (see e.g.\ \citet{StoyStoy-94}).  

Both estimators can easily be evaluated using a spherical or cylindrical coordinate system, respectively. However, unlike in the two-dimensional case, it is impossible to partition the unit sphere into equally sized cones or cylinders pointing to different directions. In practice, the directional $K$-functions can be evaluated for a set of directions evenly distributed on the unit sphere. Approaches for deriving such sets of directions are discussed in \cite{Alt11}. If possible, the choice of the number of directions should be based on prior knowledge on the main directions of anisotropy. 


While the classical summary statistics for point processes, e.g.\ Ripley's $K$-function, 
depend on one parameter, the summary statistics introduced above depend on two parameters which makes the investigations more challenging.
For the cone, it seems natural to fix the parameter $\theta$ in advance such that the conical $K$-function only depends on the parameter $r_\text{cn}$. In practice, $\theta$ should be chosen depending on the number of directions to be investigated and the intensity of the point pattern. In \cite{Redenbachetal-09} an angle of $\theta=\frac \pi 4$ was chosen when considering only coordinate directions. For larger sets of directions, $\theta$ should be reduced to avoid overlap of the cones for different directions. Additionally, the angle should be large enough to observe a reasonable number of points within the cones.  

\begin{figure}[t]
\centering
\includegraphics[width=10cm]{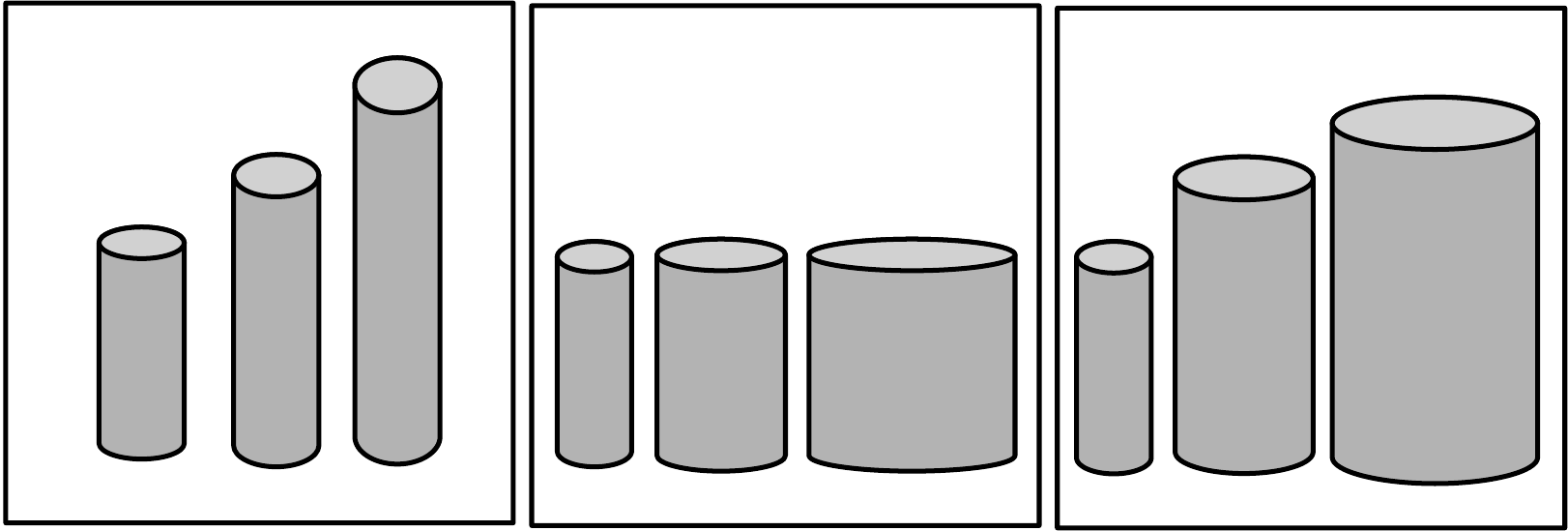}
\caption{Three ways of expanding a cylinder when detecting the anisotropy in the point patterns: expanding $h$ given a fixed $r_\text{cl}$ (left panel), expanding $r_\text{cl}$ given a fixed $h$ (middle panel), or expanding both $r_\text{cl}$ and $h$ (right panel).}
\label{fig3:cylinders}
\end{figure}

For the cylinder, the situation is more complicated as there are three ways to expand a cylinder (see Figure~\ref{fig3:cylinders}) depending on the two parameters $r_\text{cl}$ and $h$. A priori, none of these methods seems more natural than the other. In \cite{RSDRMN-15}, the height of the cylinder was fixed while expanding its radius.
  In the present study, we are interested in a comparison of the cylindrical and the conical $K$-function. Hence, the expansion scenario should be chosen such that both functions behave similarly in some sense. Two possible approaches are discussed in the following section.
 
\section{Choice of parametrization}\label{sec:volume}


\subsection{Equal volume}

The first parametrization is based on the fact that sets of equal volume will contain a similar number of points. Hence, we suggest to parametrize the functions such that the volumes of the cone and the cylinder are equal. The details are as follows.


Recall that $r_{\text{cn}}$ and $r_{\text{cl}}$ refer to the radius of the cone (or the radius of the circumscribed sphere),
 and the radius of the cylinder, respectively (see Figure~\ref{fig2:strucElmnt}). 
Knowing that the volume of a cylinder, a cone, and a spherical cap are, respectively, given by 
\begin{equation*}
V_{\text{cl}} = \pi r_{\text{cl}}^2{2h},\,  V_{\text{cone}} = \frac{1}{3}\pi r_{\text{cl}}^2{h}, V_{\text{cap}} = \frac{\pi d^2}{3}(3r_{\text{cn}} - d)
\end{equation*}
where $d = r_{\text{cn}} - h$ is the height of the cap, the volume of the double cone (used as the structuring element of the conical $K$-function) is given by 
\begin{align*}
	V_{\text{cn}} & = 2[V_{\text{cone}} + V_{\text{cap}}]\\
  	& = 2[\frac{1}{3}\pi r_{\text{cl}}^2{h} + 
  	  \frac{\pi}{3}{(r_{\text{cn}} - h)^2}(3r_{\text{cn}} - (r_{\text{cn}} - h))]\\
  	& = \frac{1}{3}\pi r_{\text{cl}}^2{2h} +  
  	\frac{2 \pi}{3}{(r_{\text{cn}} - h)^2}(2r_{\text{cn}} + h)
\end{align*}

Using the above formula, those values of $r_{\text{cn}}$, $r_{\text{cl}}$, and $h$ satisfying 
\begin{equation}\label{volume}
	r_{\text{cl}}^2{2h} = 
  	{(r_{\text{cn}} - h)^2}(2r_{\text{cn}} + h).
\end{equation}
lead us to the equation $V_{\text{cn}} = V_{\text{cl}}$, i.e., the equality of the volumes of the structuring elements of the two functions. 

Equation \eqref{volume} leaves two degrees of freedom. In practice, it can be accompanied by further constraints such as the choice of an aspect ratio for the cylinder (see below).

\subsection{Equal shape}

An alternative approach is to require that similar regions of the data are scanned in the sense that the shapes of the structuring elements are similar. This is achieved by placing the cone inside the cylinder as shown in Figure~\ref{fig2:strucElmnt}. In this case, the following equations hold:
\begin{equation}
\label{AR}
\cot(\theta)=\frac{h}{r_{\text{cl}}}
\end{equation}
and
\begin{equation}
\label{rrh}
r^2_{\text{cn}} = h^2 + r^2_\text{cl}.
\end{equation}
Following the recommendation given in \citet{plcpp-15} on using an elongated cylinder, i.e.\ where $h>r_\text{cl}$,  the right hand side of equation \eqref{AR} can be considered as an aspect ratio. It is clear that when this ratio is equal to one, no anisotropy is expected to be detected by this function. Taking an aspect ratio $\cot(\theta)=a > 1$ and using equations~\eqref{AR} and \eqref{rrh} results in
\begin{equation}
\label{rh}
h = ar_{\text{cl}}
\end{equation}
and 
\begin{equation}
\label{rr}
r_{\text{cn}} =  r_{\text{cl}}\sqrt{a^2 + 1}
\end{equation}
which provides us with an alternative relationship between the three parameters $r_{\text{cn}}$, $r_{\text{cl}}$ and $h$. In the following, we will use the parametrization based on equations~\eqref{rh} and \eqref{rr}.

\subsection{Isotropy test}\label{sec:isotropytest}

\citet{Redenbachetal-09} introduced a non-parametric method to detect anisotropies in the point patterns as follows. Assuming isotropy in the $xy$-plane and knowing that the anisotropy is directed along the $z$-axis, the isotropy test for $m$ replicated point patterns is based on the statistics given by 
\begin{equation*}
T_{xy,i}= \int_{r_1}^{r_2} | \hat{S}_{x,i}(r)-\hat{S}_{y,i}(r)| dr
\end{equation*}
and 
\begin{equation*}
T_{z,i}= \min  \left( \int_{r_1}^{r_2} | \hat{S}_{x,i}(r)-\hat{S}_{z,i}(r)| dr, \int_{r_1}^{r_2} | \hat{S}_{y,i}(r)-\hat{S}_{z,i}(r)| dr \right)
\end{equation*}
where $[r_1, r_2]$ is a given interval, and $\hat S_x$, $\hat S_y$, and $\hat S_z$ are estimates of a summary statistic (here, either the conical or the cylindrical $K$-function) in the directions of the $x$-, $y$-, and $z$-axis, respectively. Here, $r_\text{cl}$ or $r_\text{cn}$ are chosen as the integration variable while the remaining parameters $h$ and $\theta$ are chosen by any of the approaches discussed above.

In case of isotropy, these three estimates should behave similarly, while $\hat{S}_z$ should be clearly different from $\hat{S}_x$ and $\hat{S}_y$ if the anisotropy is directed along the $z$-axis. Hence, the null hypothesis of isotropy will be rejected at significance level $\alpha$ if the value of $T_{z,i}$ corresponding to the $i$-th point pattern is larger than $100(1-\alpha)\%$ of the estimated $T_{xy, i}$ values. The performance of the test is evaluated using its power, estimated by the average number of times the null hypothesis is rejected in $1000$ repetitions of the test. Note that the values of $r_1$ and $r_2$ should be chosen depending on the type of anisotropy. We fix $r_1=0$ and will investigate the effect of different choices of $r_2$ on the power of the test (see also \citet{Redenbachetal-09}). 

\begin{figure}[t]
\centering
\includegraphics[width=4.5cm]{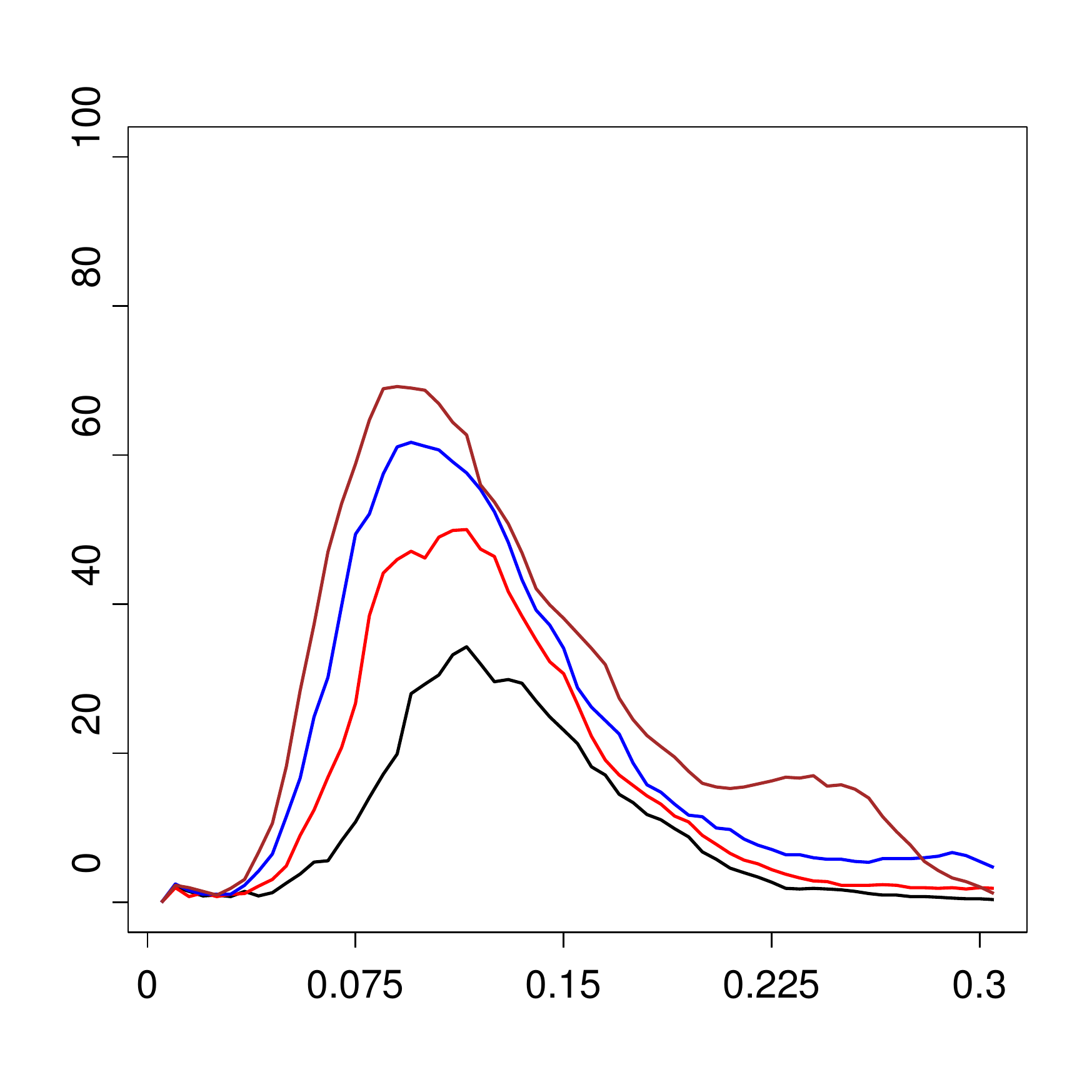}
\includegraphics[width=4.5cm]{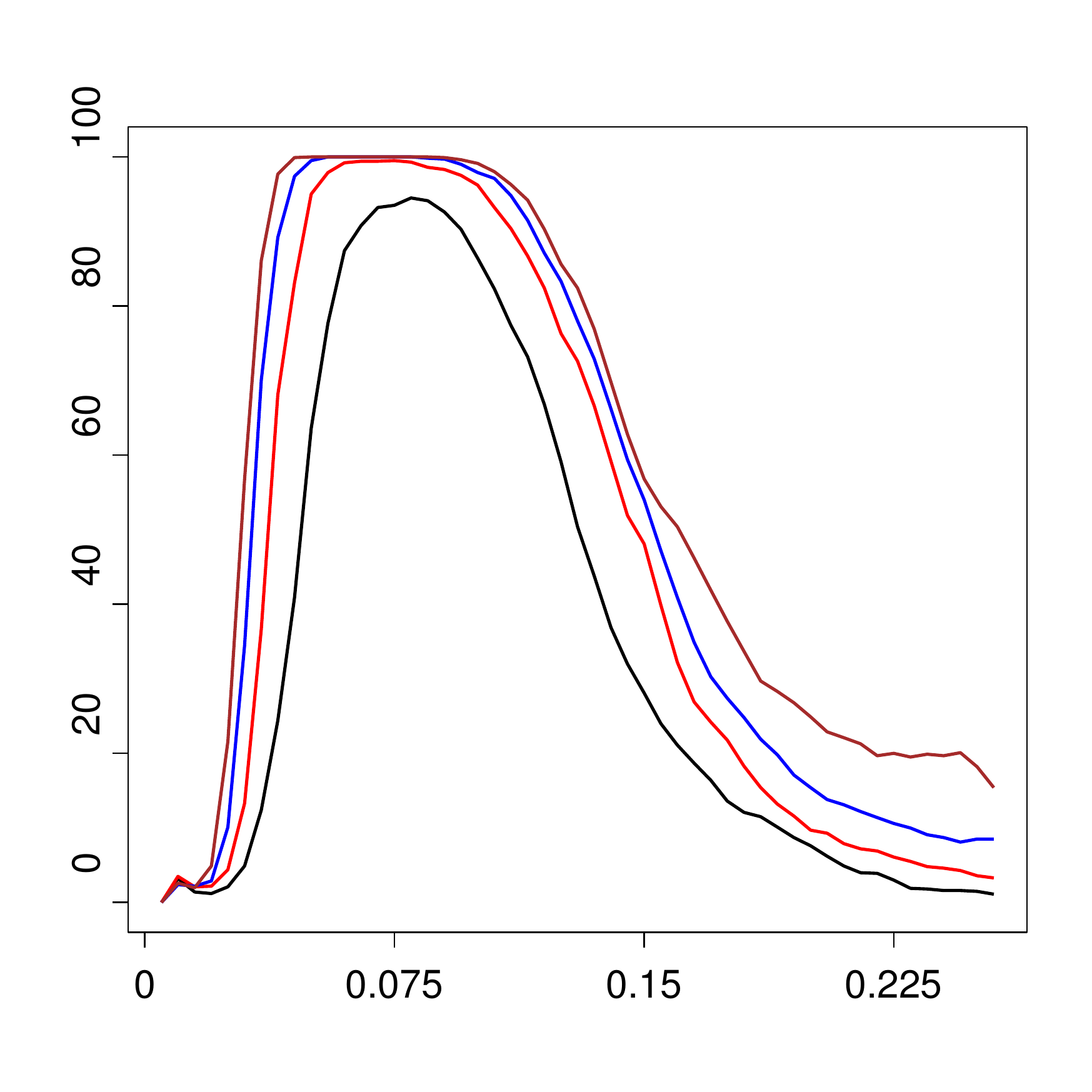}\\
\includegraphics[width=4.5cm]{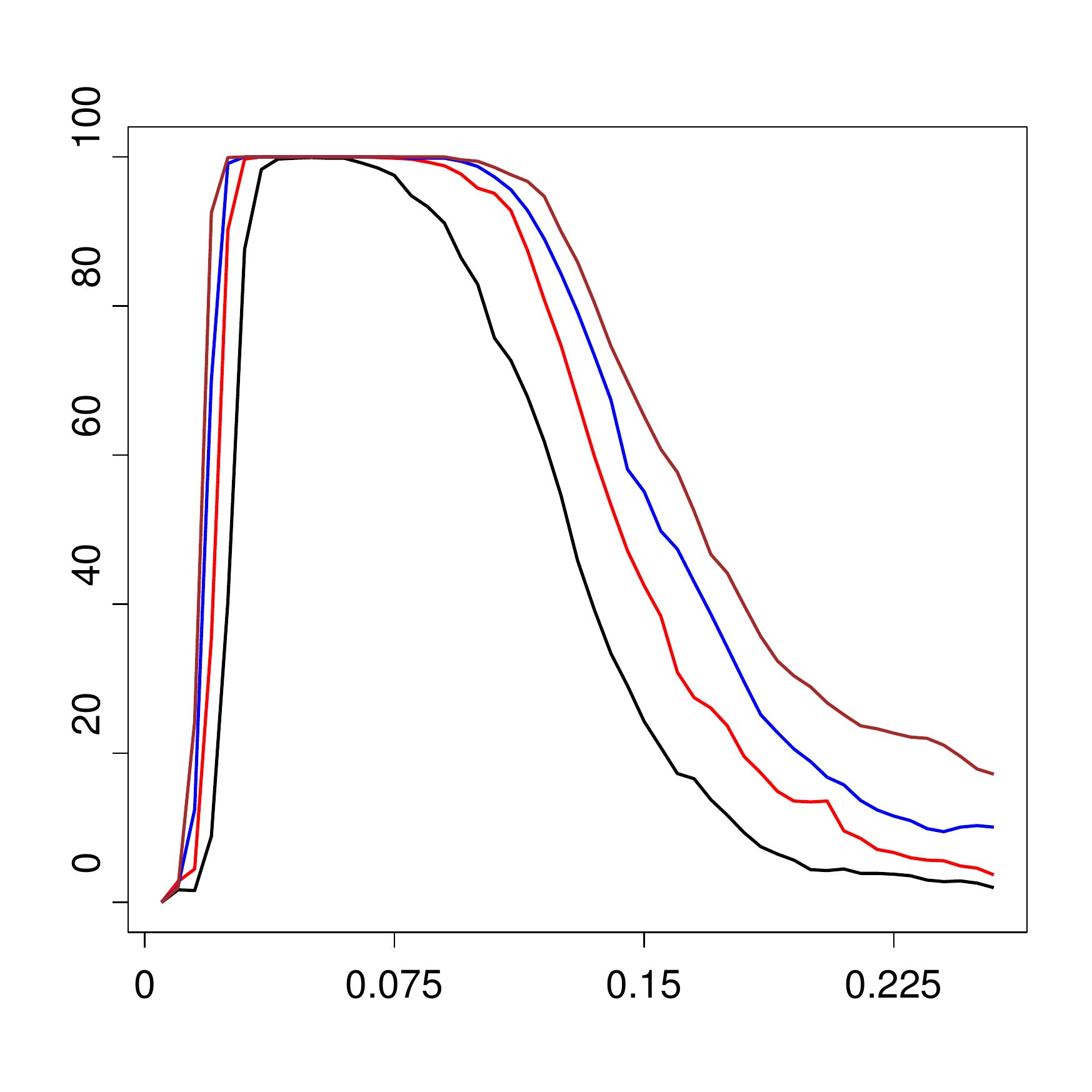}
\includegraphics[width=4.5cm]{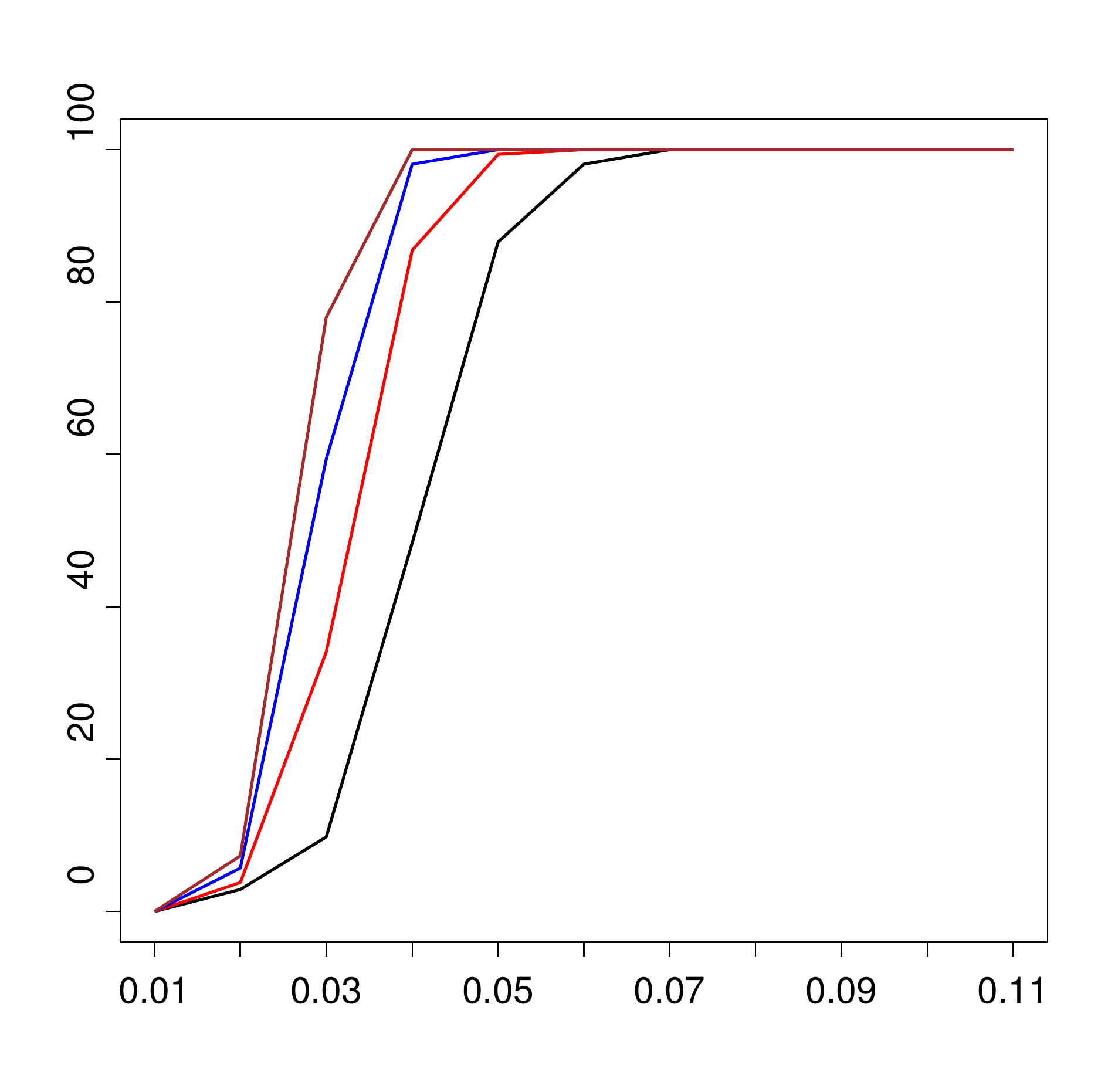}
\caption{The power of the isotropy test at level 5\% versus the parameter $r_2$ for the realizations of the PLCPP model 
 with (from top left to bottom right) very low, low, medium, and high degree of linearity based on the cylindrical $K$-function. 
The curves from bottom (black) to top (brown) correspond to the aspect ratios from $a = 1.5$ to $a = 3$ with an increment of size $0.5$.
}
\label{fig4:Power}
\end{figure}

When using the equations obtained in the above sections, one should also decide on an appropriate aspect ratio $a$. Figure ~\ref{fig4:Power} shows plots of the power of the isotropy test at level 5\% versus the parameter $r_2$ for the cylindrical $K$-function (and the corresponding $r_2$ for the conical $K$-function obtained using \eqref{rr})
 for the aspect ratios $a$ from $1.5$ to $3$ with an increment of size $0.5$, based on $m = 1000$ realizations under the PLCPP model introduced in Section~\ref{sec:data-plcpp}. 
 
The results indicate that the use of longer cylinders results in larger powers of the isotropy tests. This supports the recommendation given in \citet{plcpp-15} on using an elongated cylinder. In each plot, the maximum is obtained for approximately the same $r_2$ value, no matter which $h$ is chosen. For higher degrees of linearity, the power of the test is higher in general. Furthermore, it is less sensitive to the choice of $r_2$. 
 
\section{Application}
\label{sec:examples}


Even though the findings presented in the previous section suggest using a cylinder as long as possible, we have chosen an aspect ratio of $a=2$ for the subsequent analyses. The reasons are as follows: When using a very long cylinder, serious edge effects may occur already for small values of $r_{\text{cl}}$ resulting in poor estimates of the cylindrical $K$-function. In addition, increasing the length of the cylinder would mean to reduce the angle used for the cone. As we already mentioned, one should make sure that the cone is not too narrow as in this case it will only contain very few points. 
\begin{figure}[p]
\centering
\includegraphics[width=5cm]{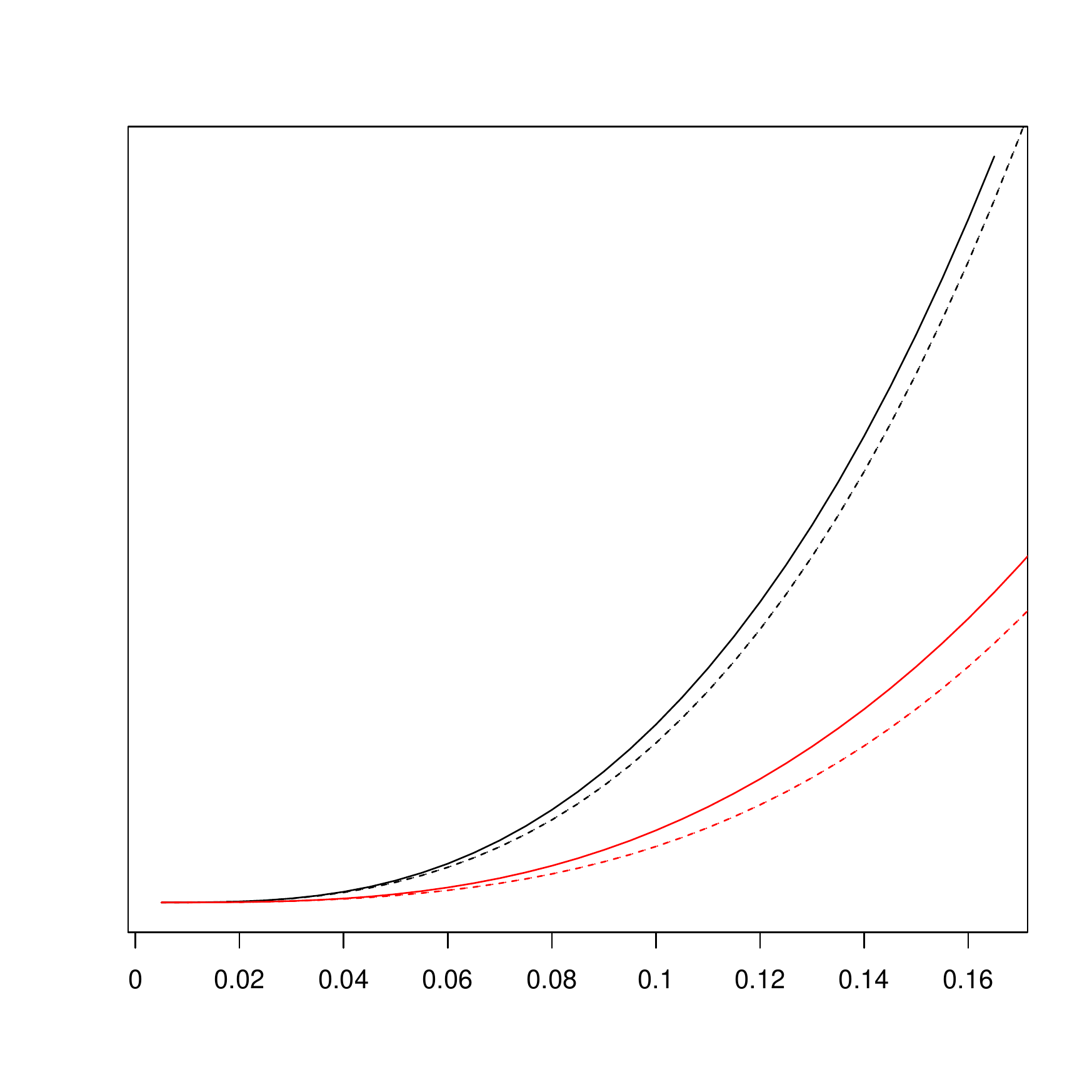}
\includegraphics[width=5cm]{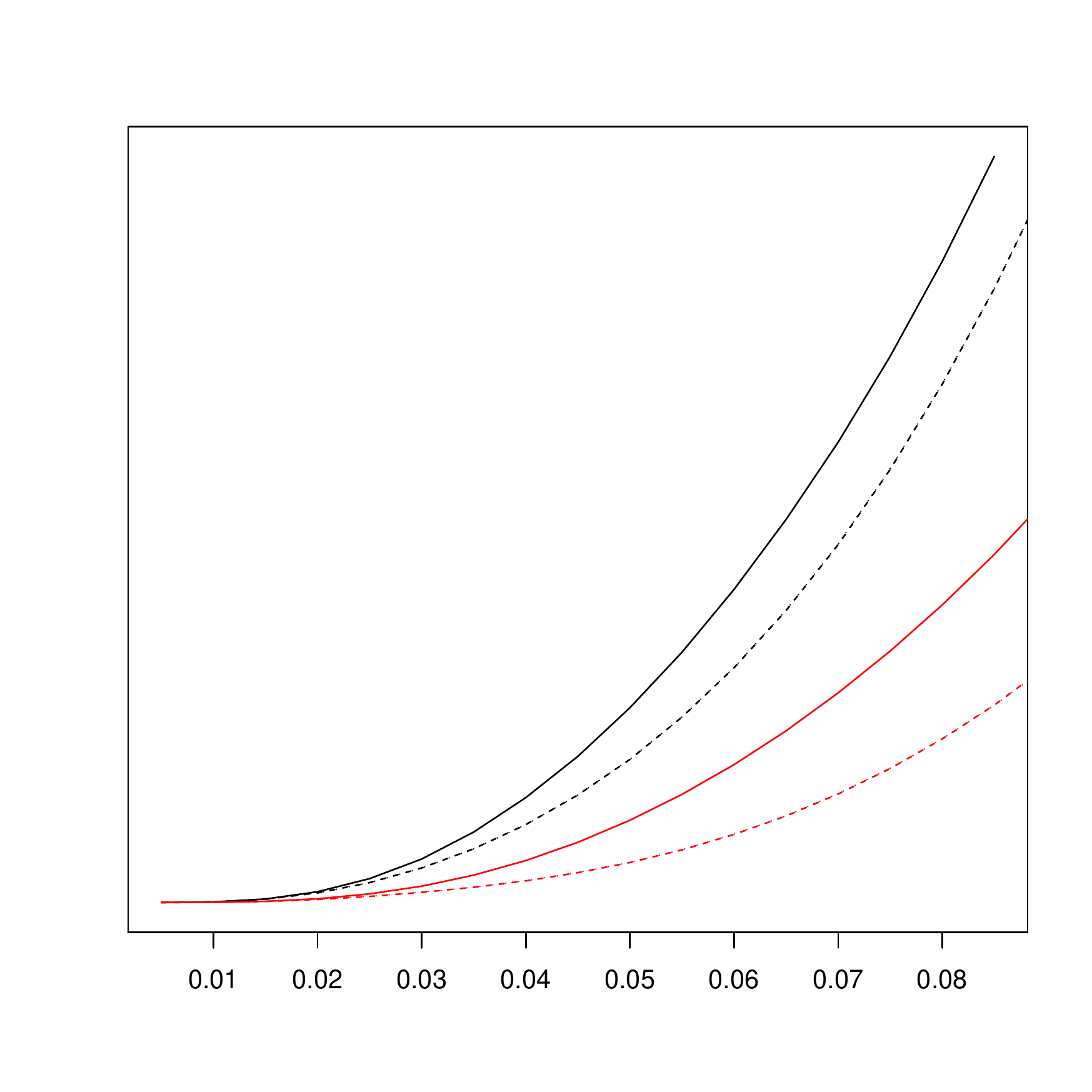}\\
\includegraphics[width=5cm]{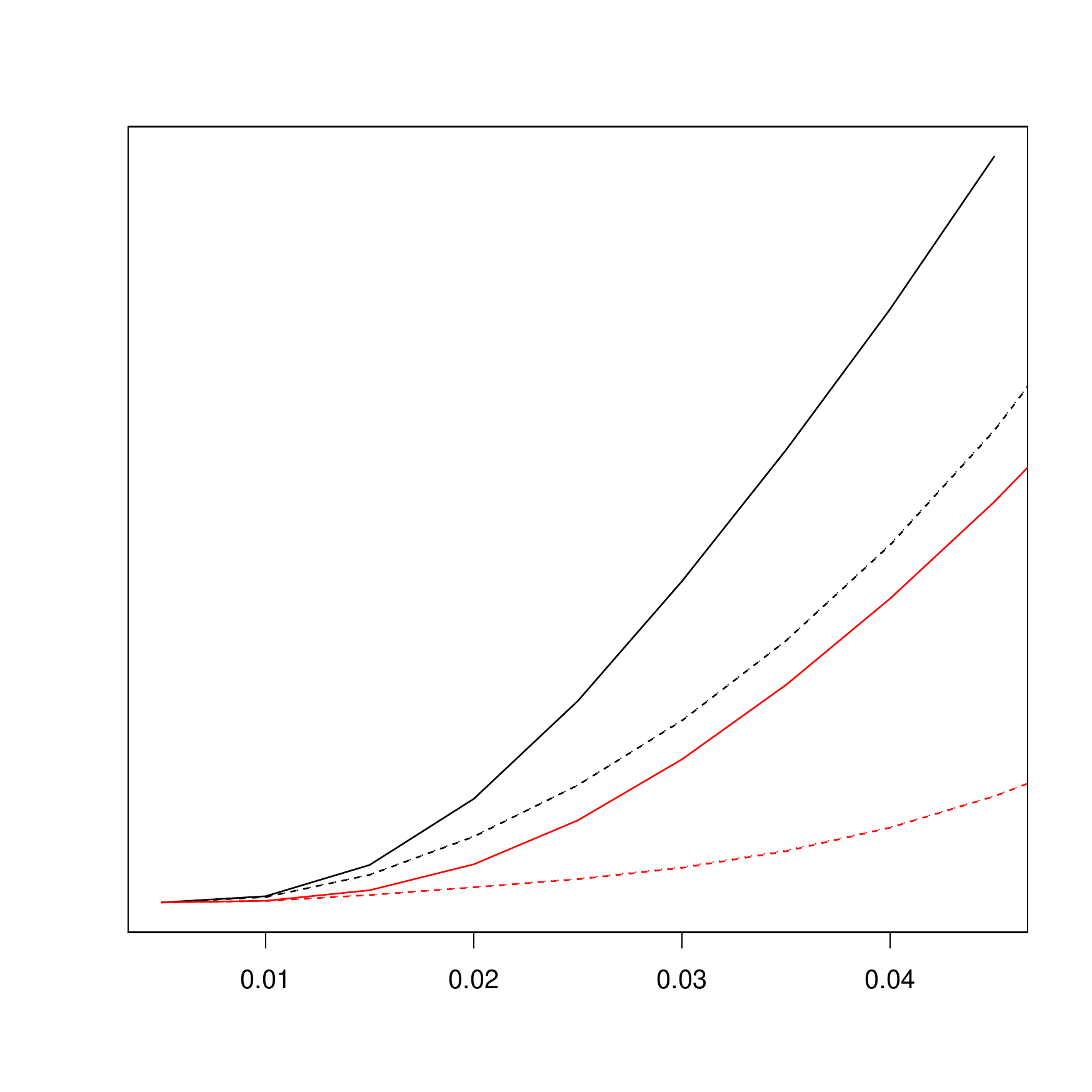}
\includegraphics[width=5cm]{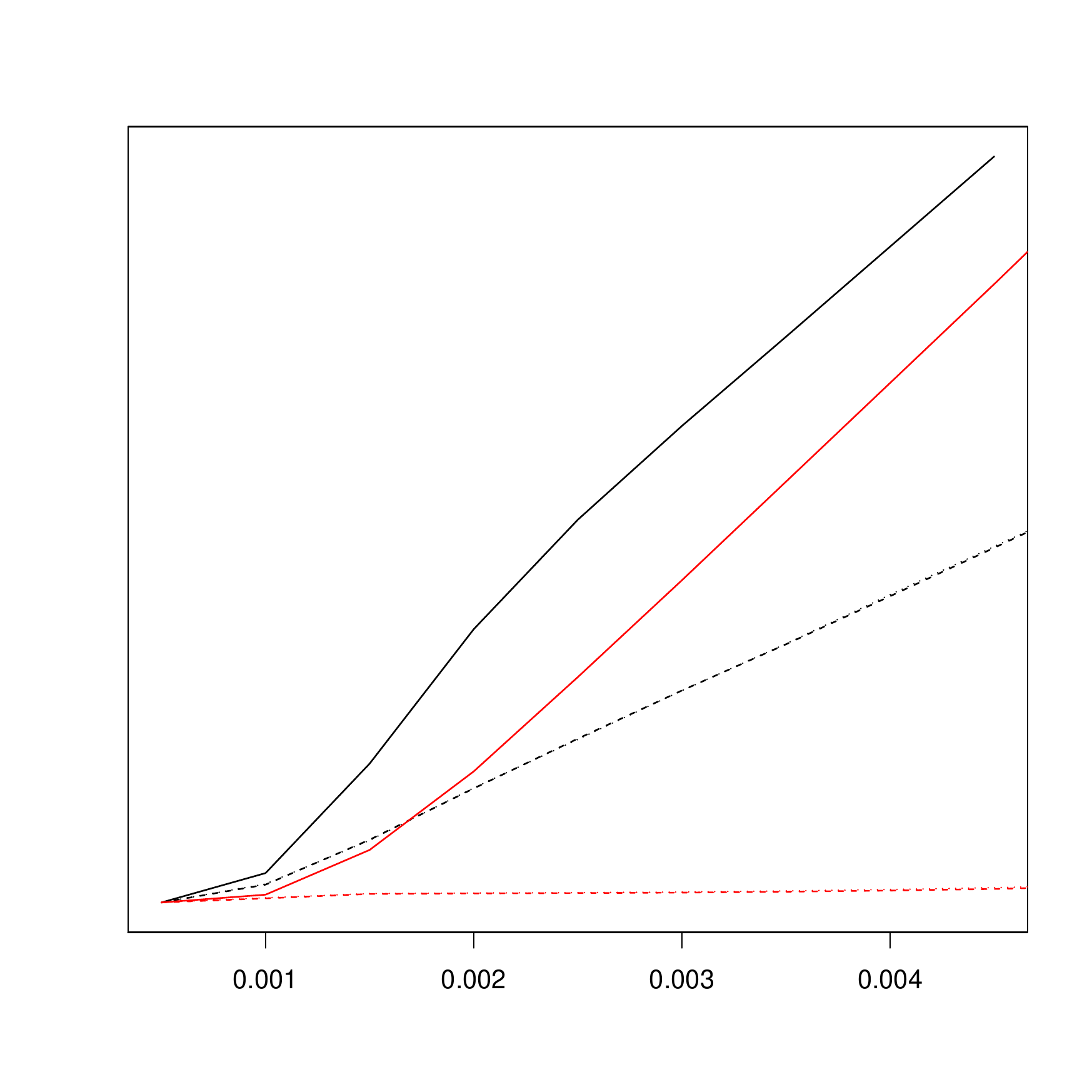}\\
\includegraphics[width=4cm]{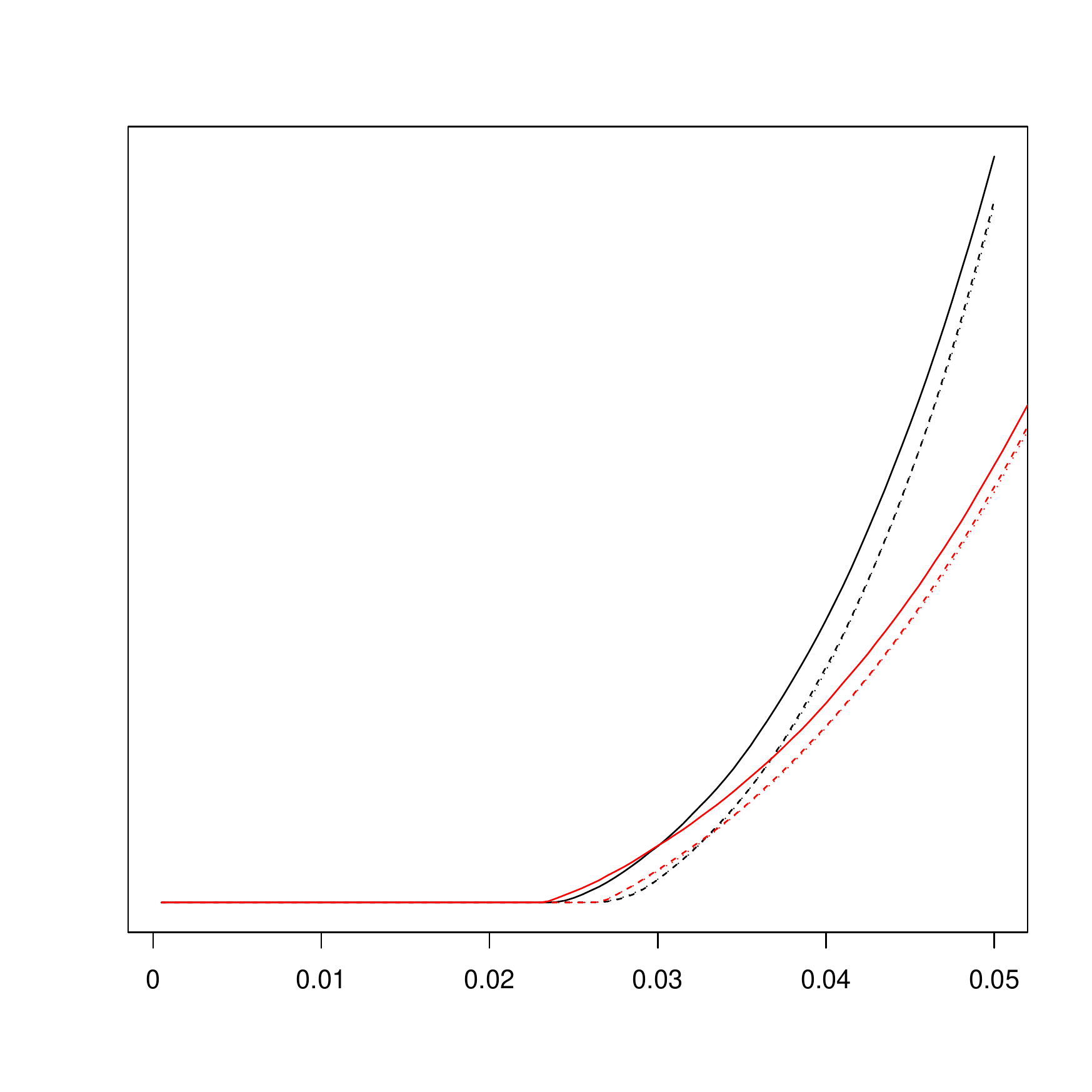}
\includegraphics[width=4cm]{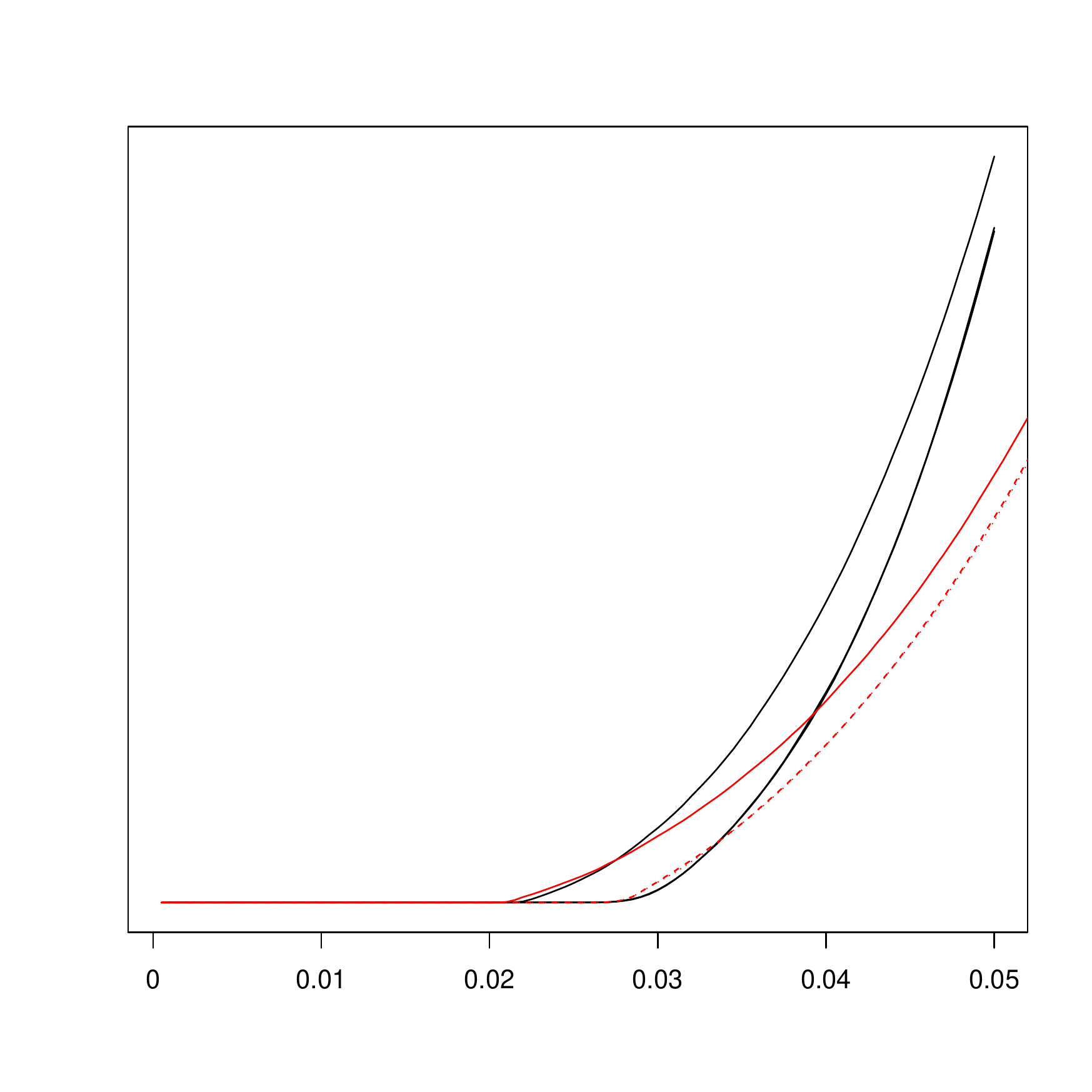}
\includegraphics[width=4cm]{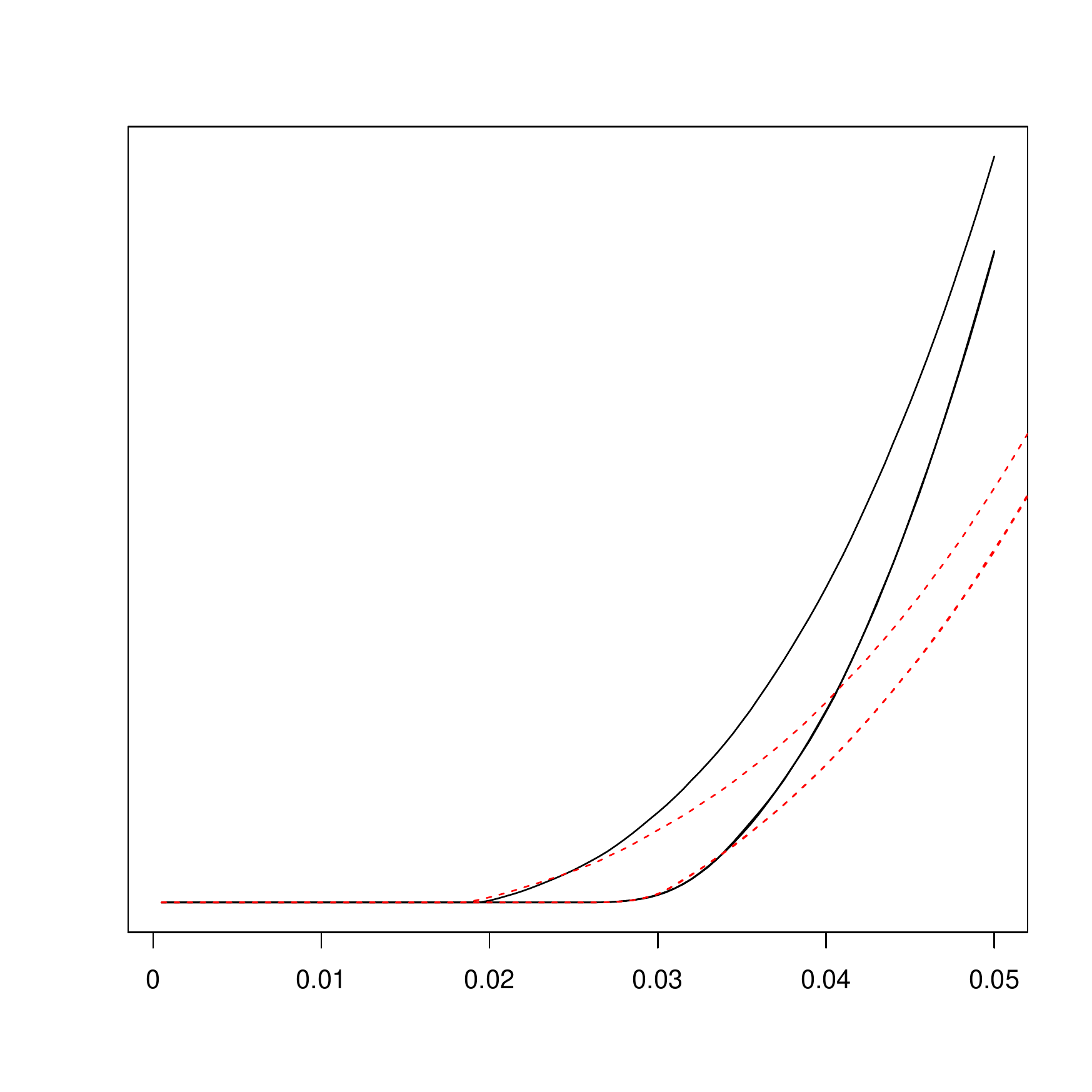}\\
\includegraphics[width=4cm]{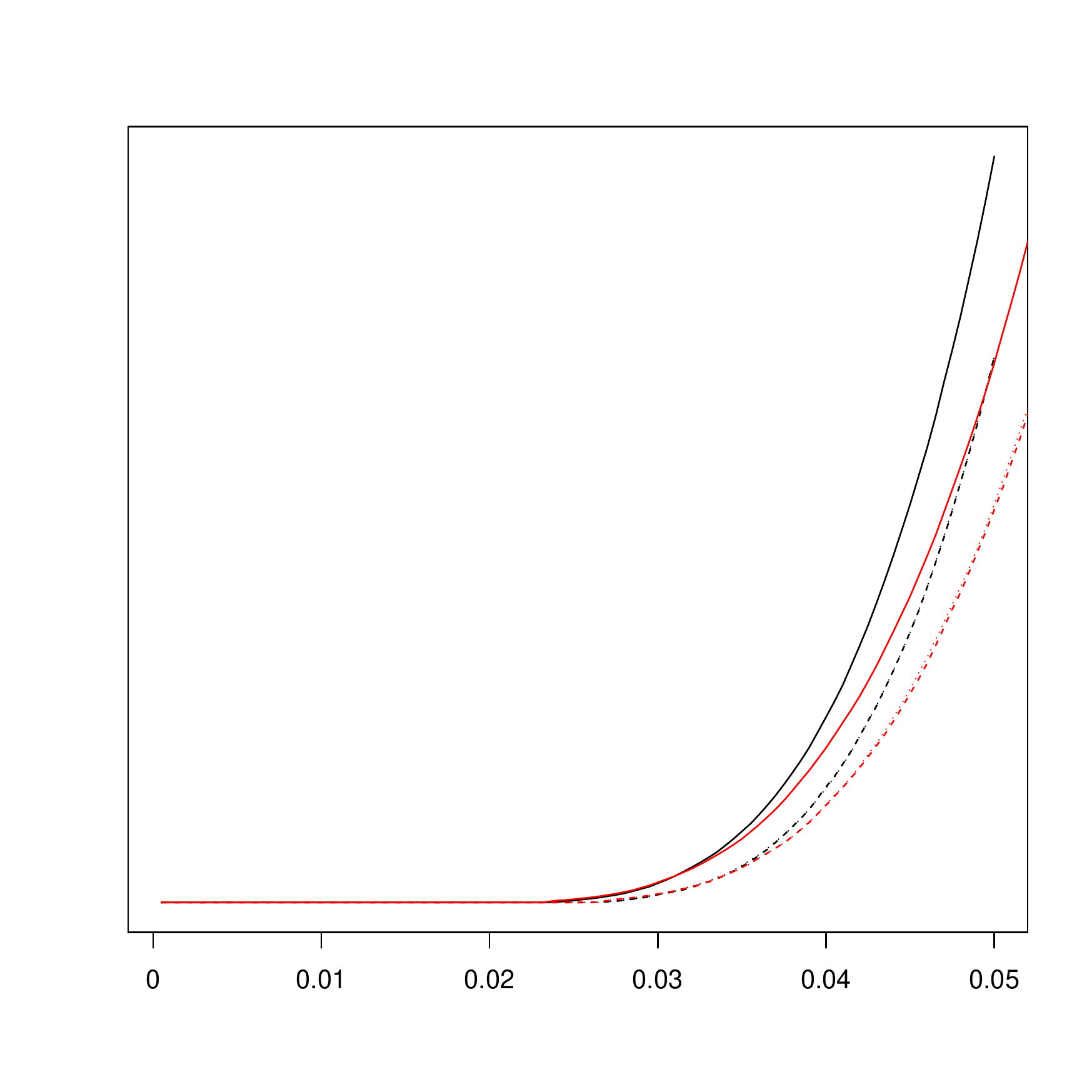}
\includegraphics[width=4cm]{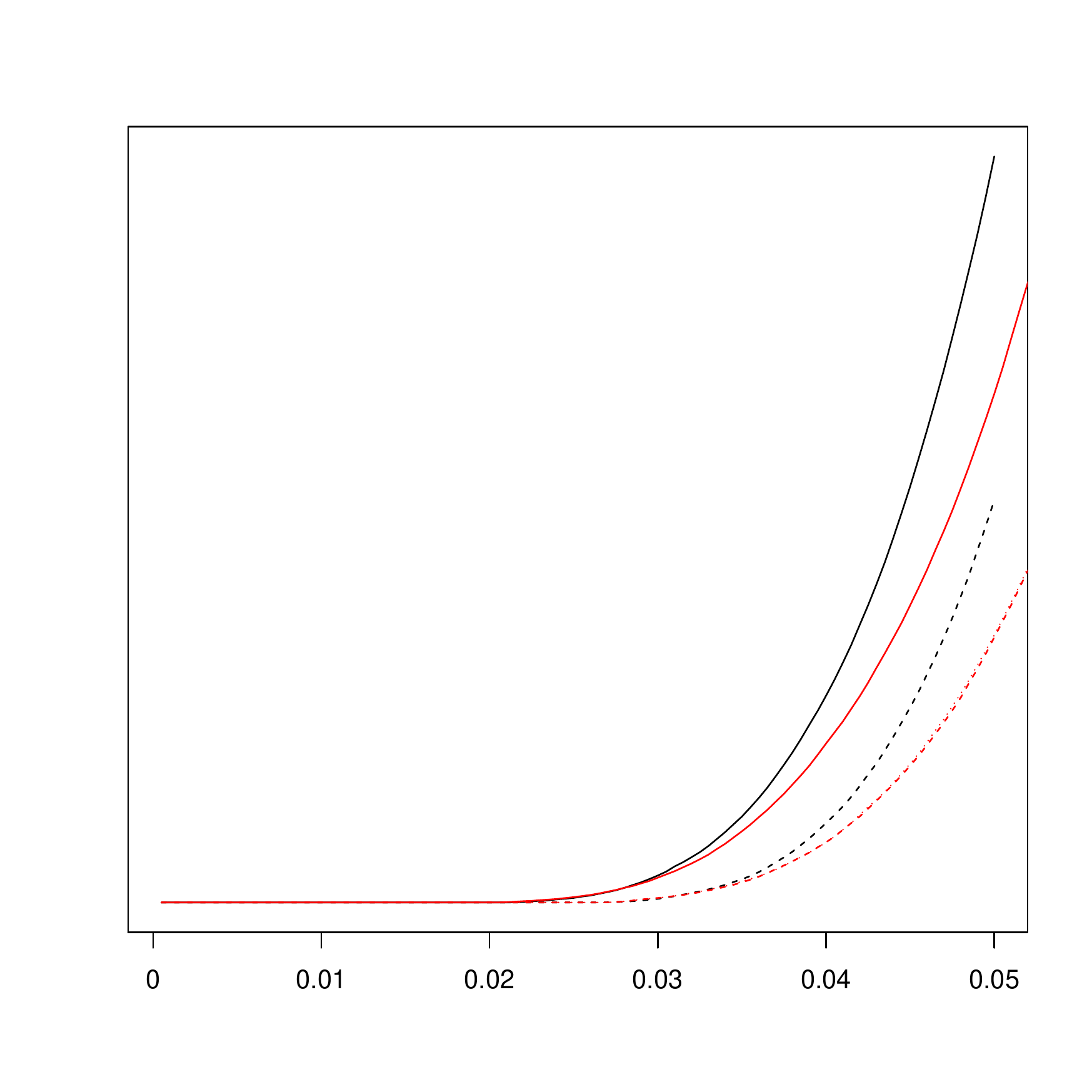}
\includegraphics[width=4cm]{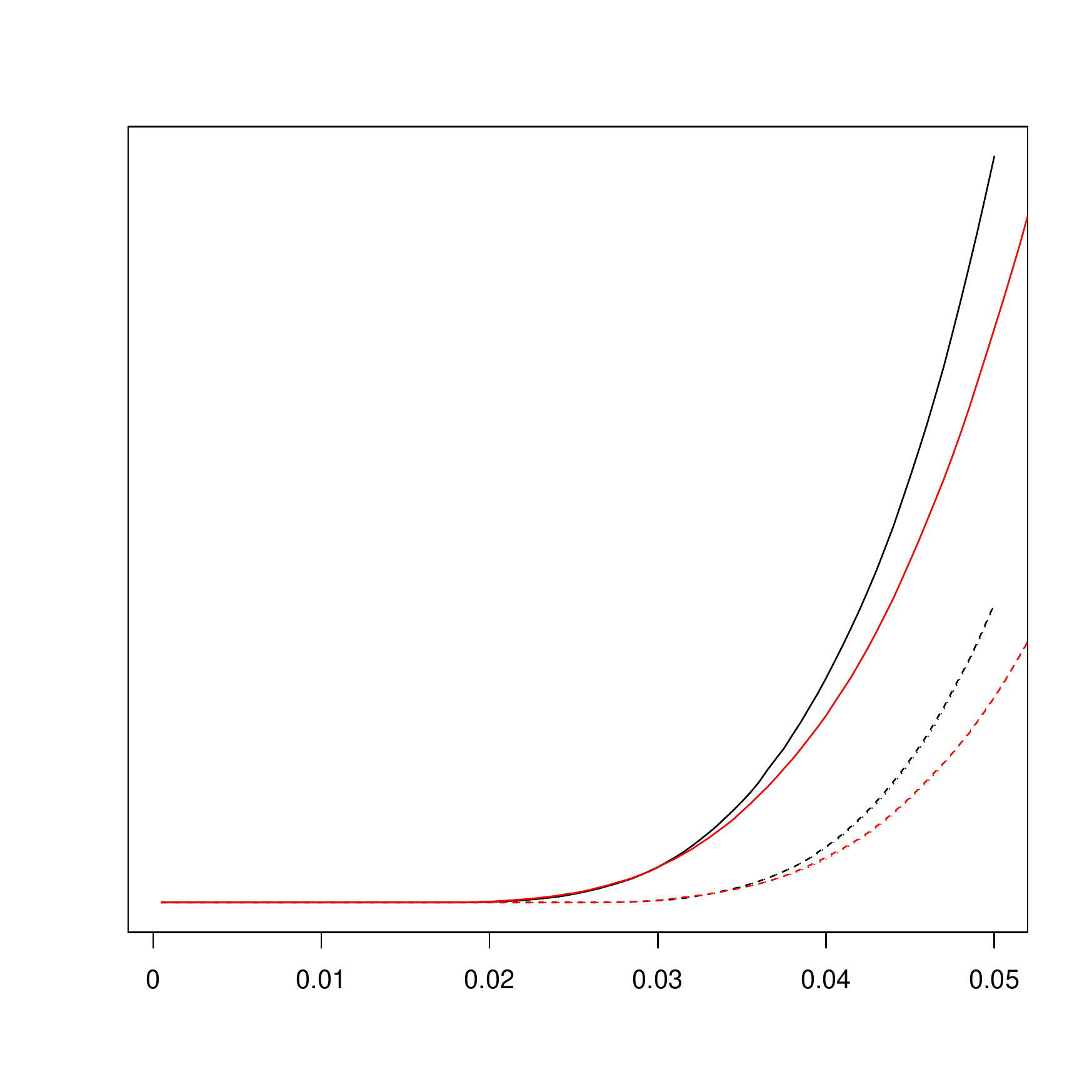}
\caption{Means of the estimated values of the cylindrical (black) and conical (red) $K$-functions 
for the realizations of the PLCPP with $\sigma = 0.04, 0.02, 0.01, 0.001$ (first four panels from top left to bottom right), Mat\'{e}rn hard-core (third row), and random packing of balls (last row), in the direction of the $z$ (solid), $x$ (dotted), and $y$ (dashed) axis. The order in the last two rows corresponds to the factors $c = 0.9, 0.8, 0.7$, from left to right, respectively. 
}
\label{fig5:MeanCylKDirK}
\end{figure}

Hence, in the applications, we chose $\theta = 0.4636476$ which is corresponding to $a = 2$, i.e.\  the case where the height of the cylinder is twice the diameter of its base. Figure~\ref{fig5:MeanCylKDirK} shows the means of the estimated values of conical and the cylindrical $K$-functions for 1000 realizations 
 of the simulated data sets introduced in Sections~\ref{sec:data-plcpp} and  \ref{sec:regular}. The mean values are obtained using the ratio estimation method described in \citet{badRep-93}. The $x$-axis of the plot shows the values for $r_\text{cl}$ (and the corresponding parameters $r_\text{cn}$ and $h$ are obtained from equations~\eqref{rh} and \eqref{rr} to get comparable scales).  
With the exception of the Mat\'{e}rn case  where the anisotropy is only weakly pronounced, both functions are able to detect the anisotropy. However, it is not easy to see which function is more sensitive to the structure of the anisotropy. Therefore, we made a comparison based on the power of the isotropy tests as follows.
\begin{figure}[p]
\centering
\includegraphics[width=4cm]{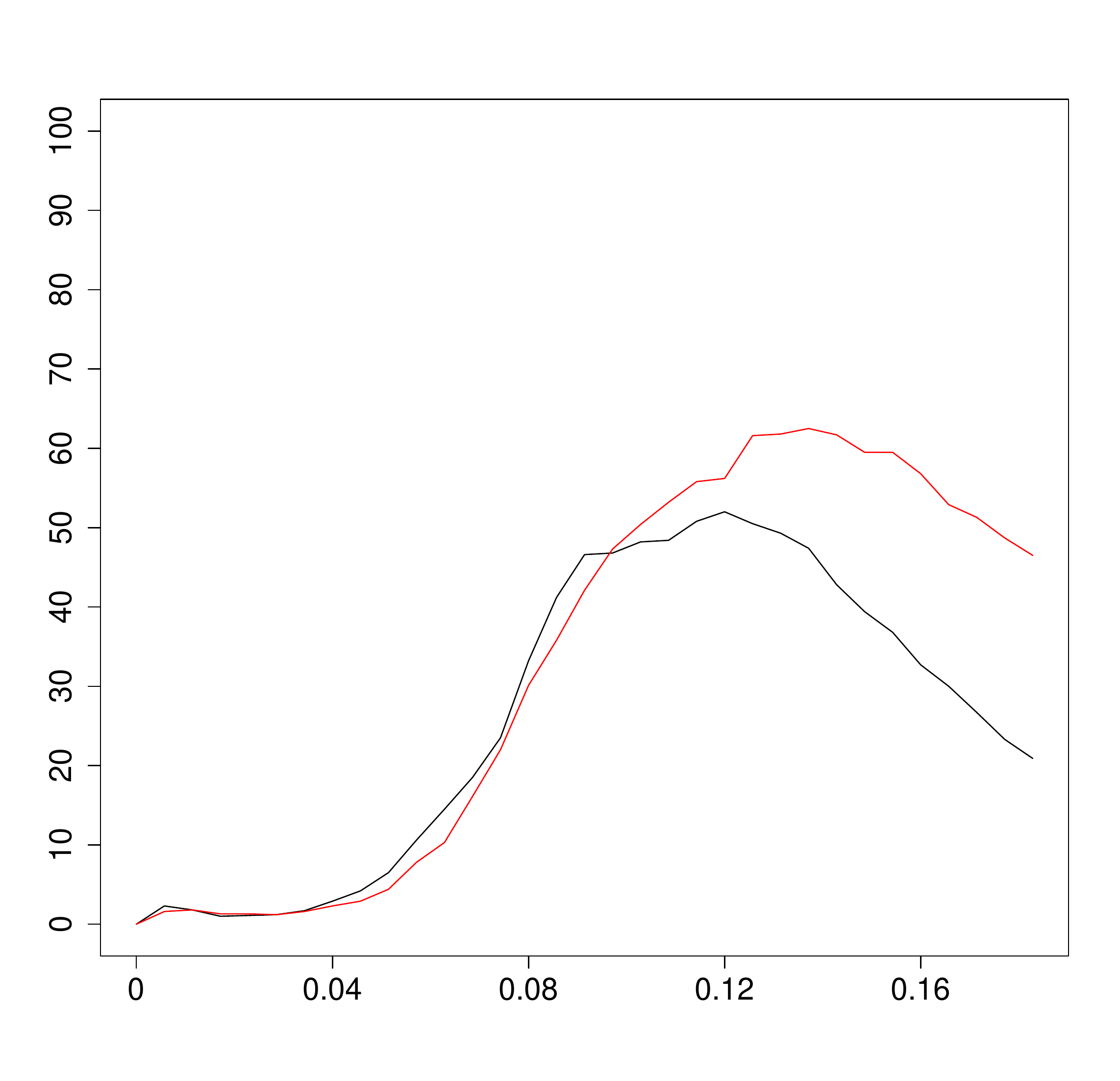}
\includegraphics[width=4cm]{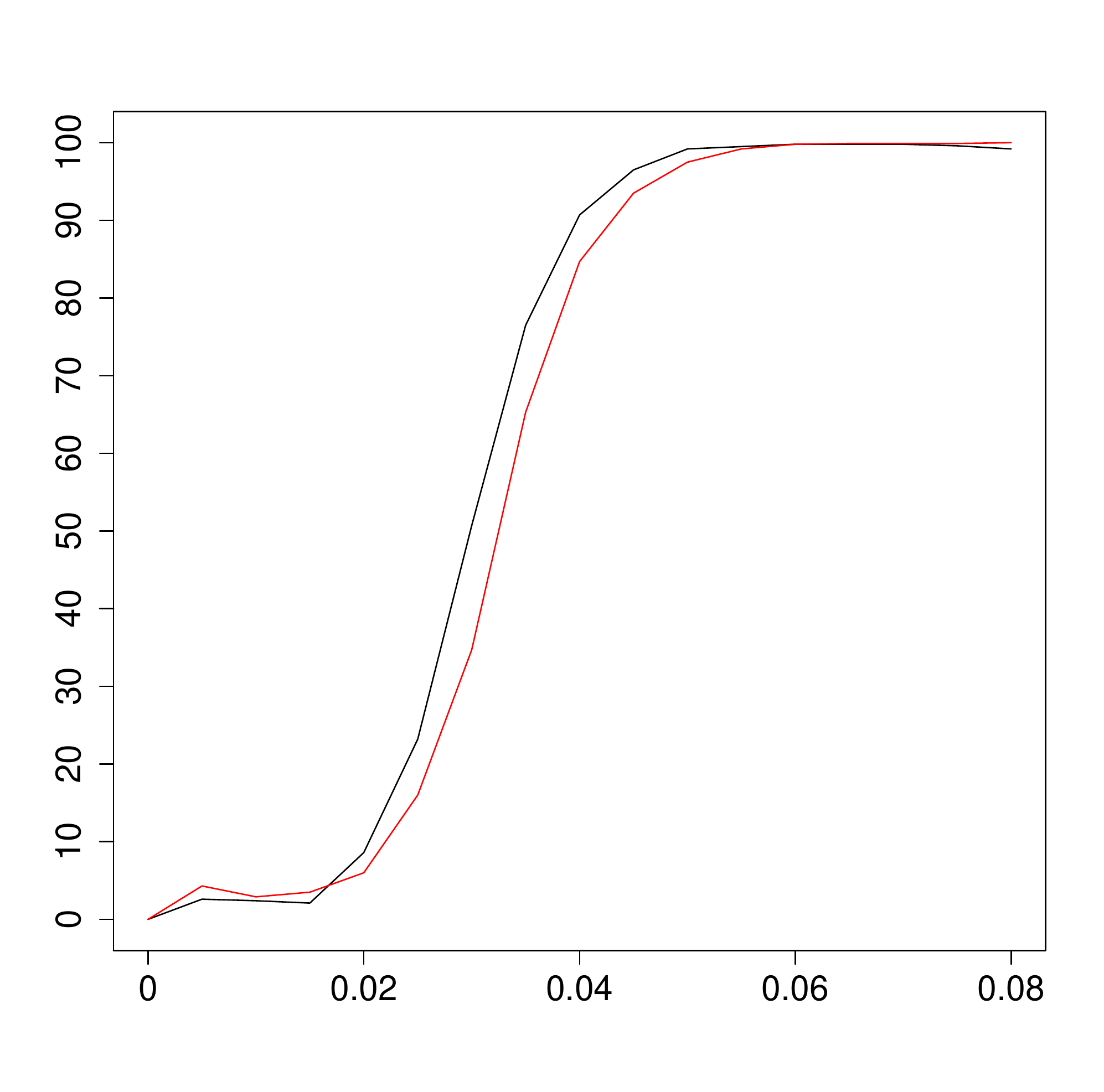}\\
\includegraphics[width=4cm]{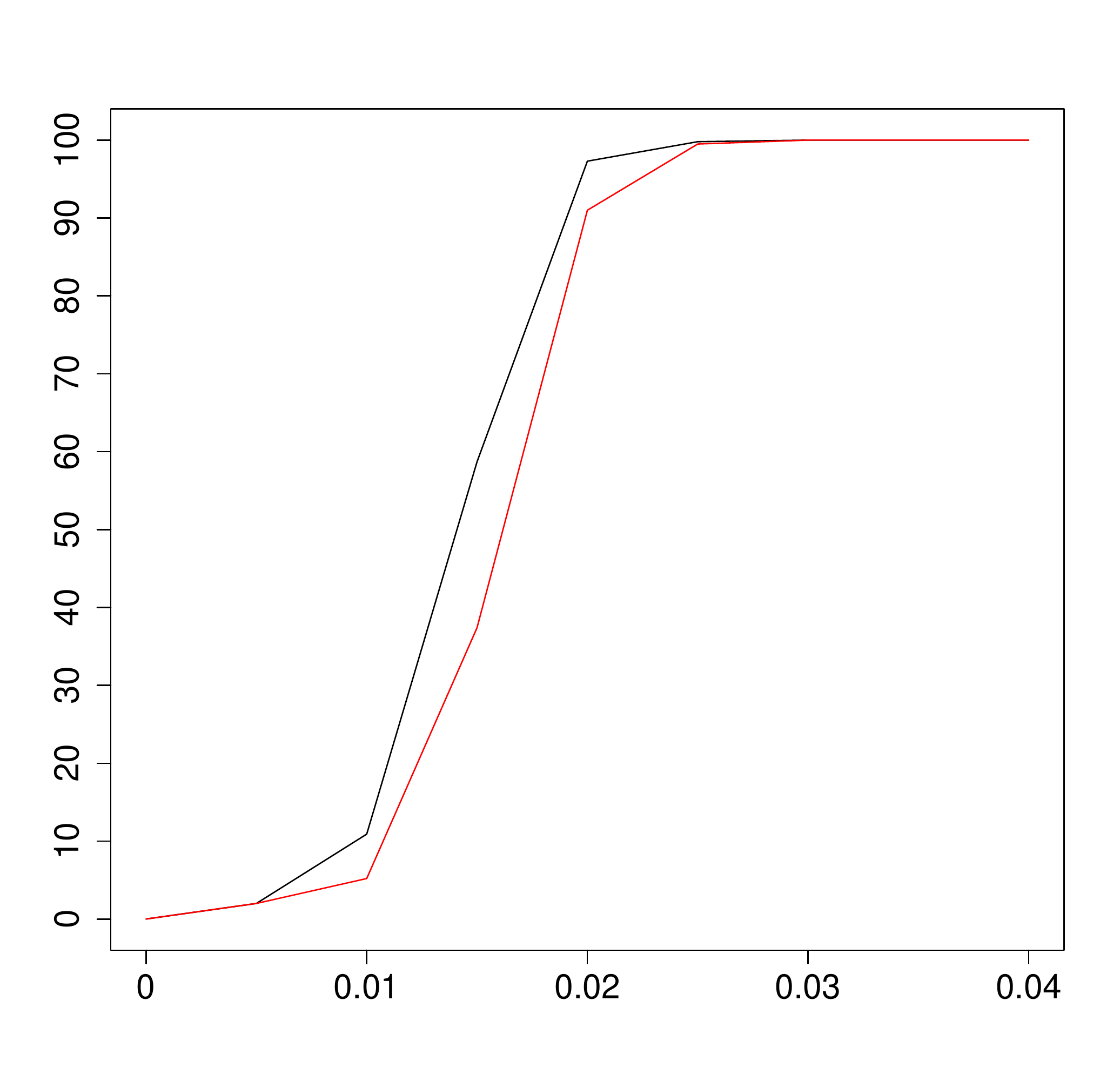}
\includegraphics[width=4cm]{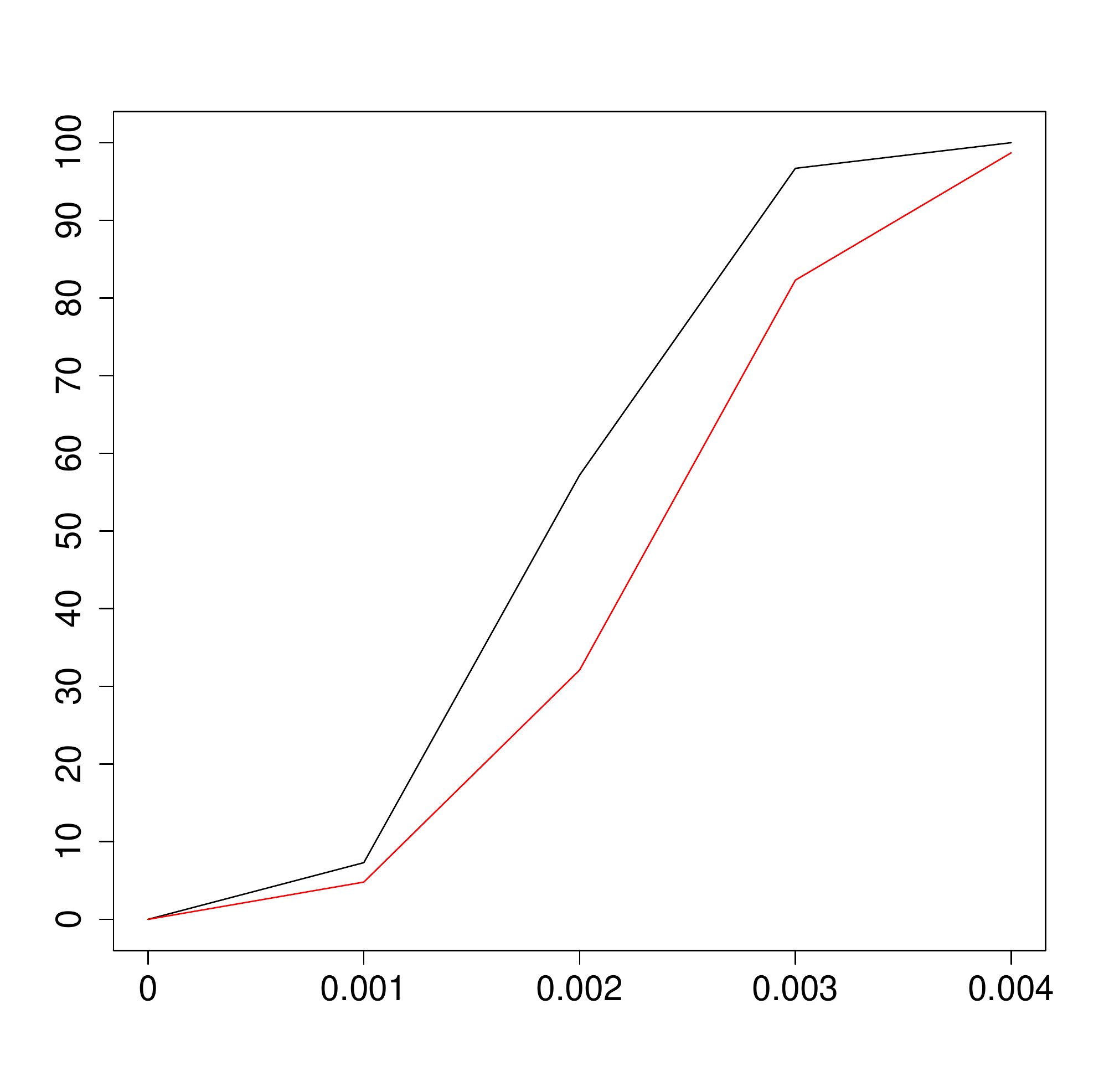}\\
\includegraphics[width=4cm]{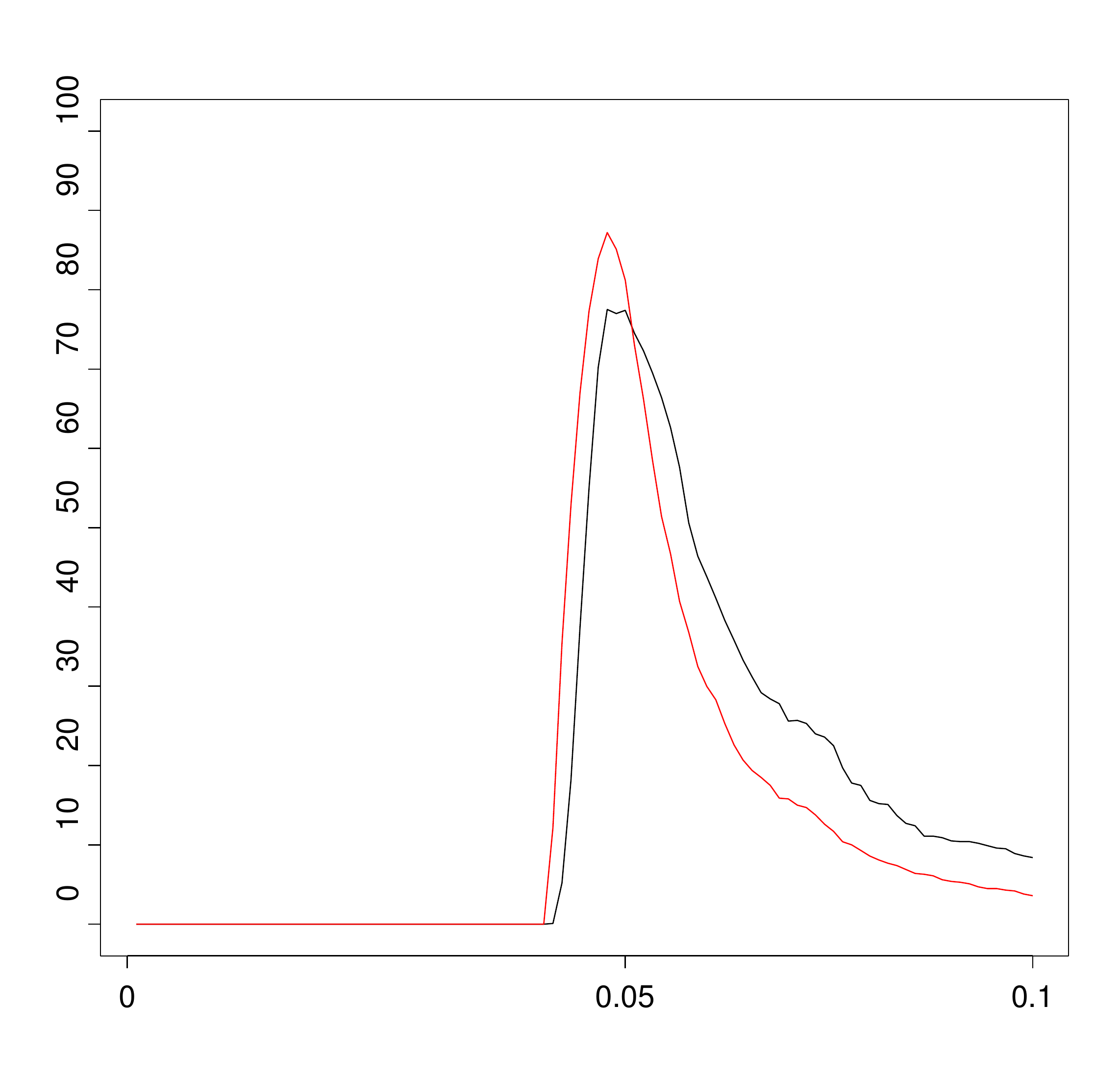}
\includegraphics[width=4cm]{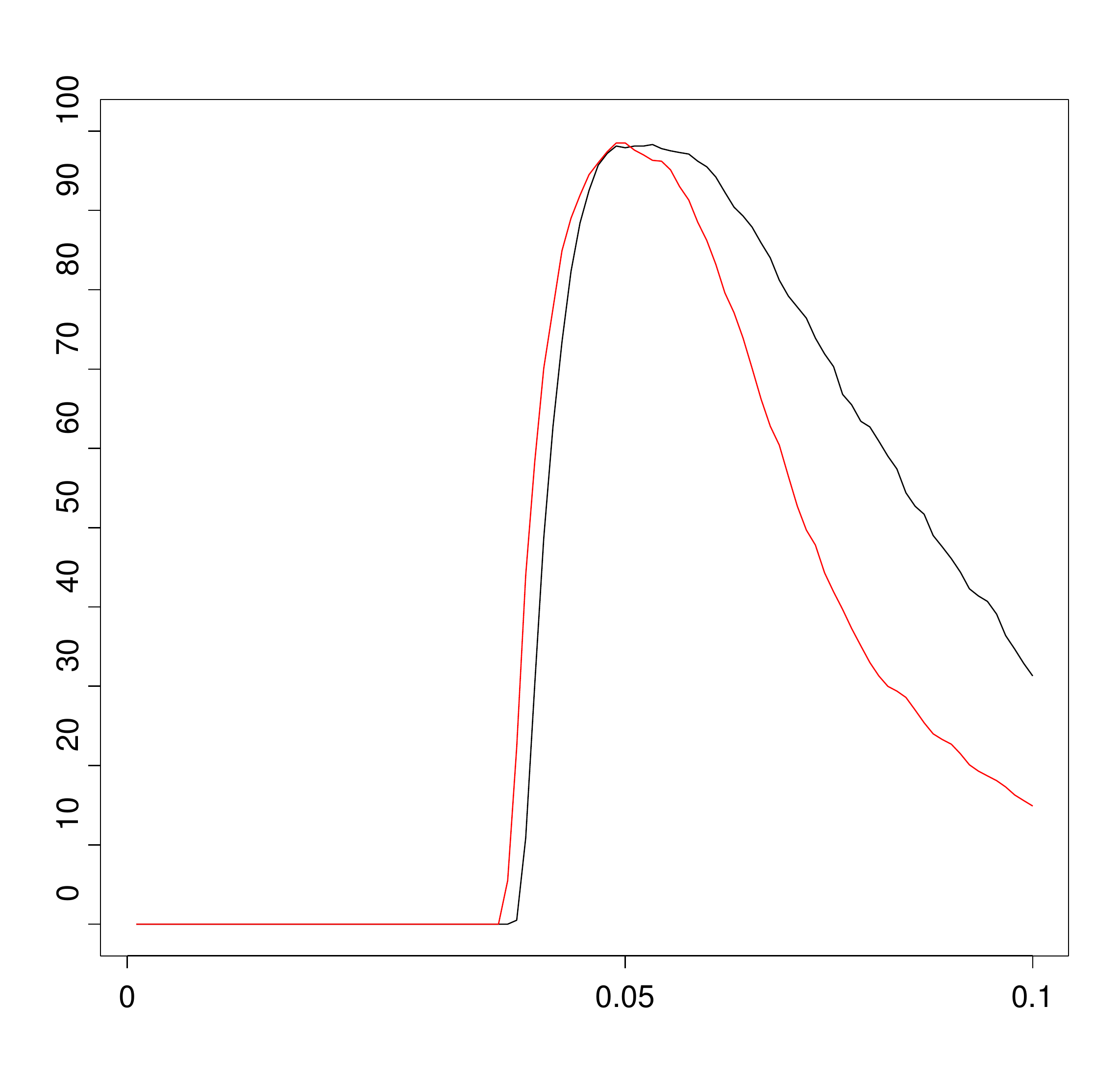}
\includegraphics[width=4cm]{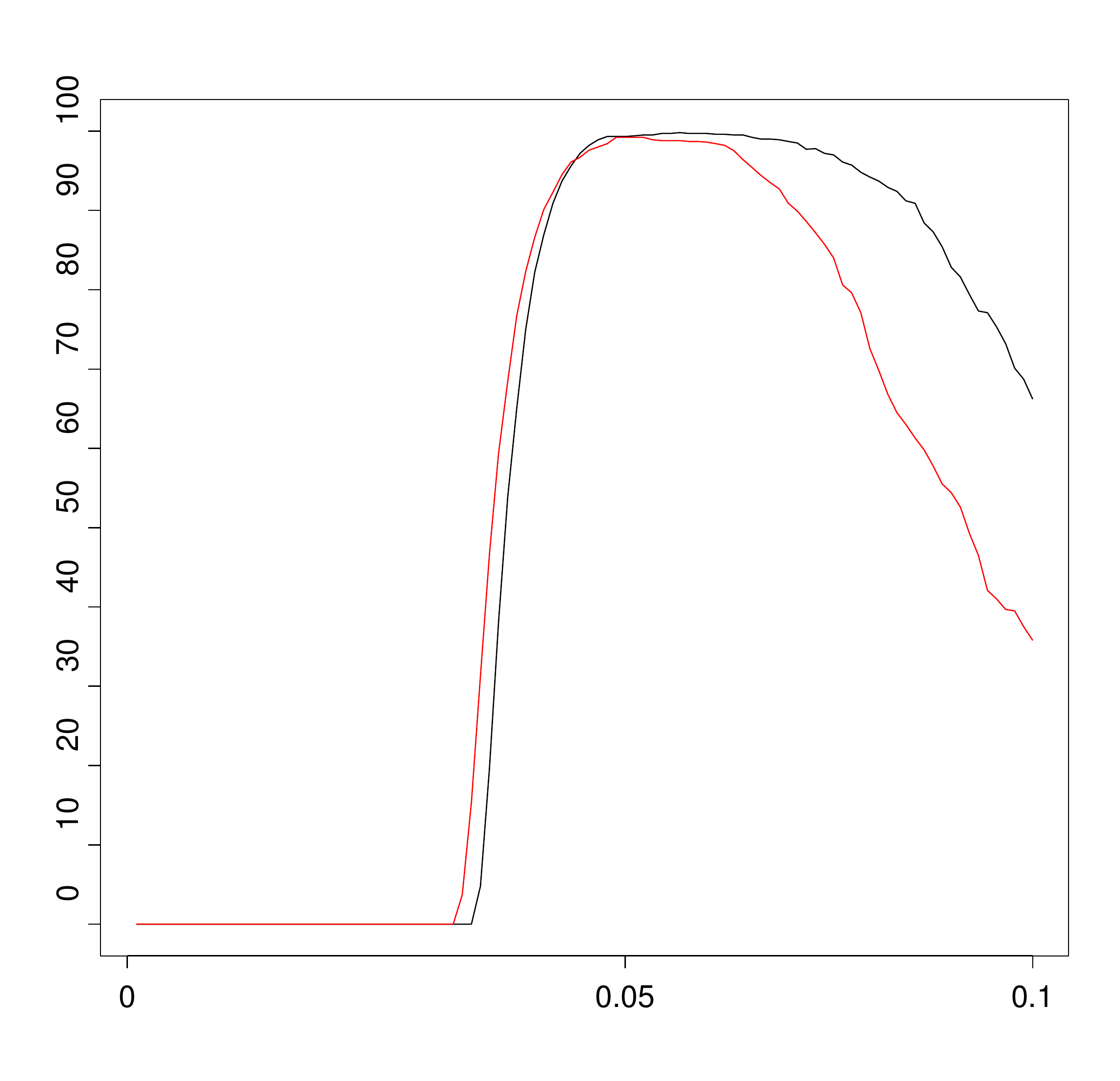}\\
\includegraphics[width=4cm]{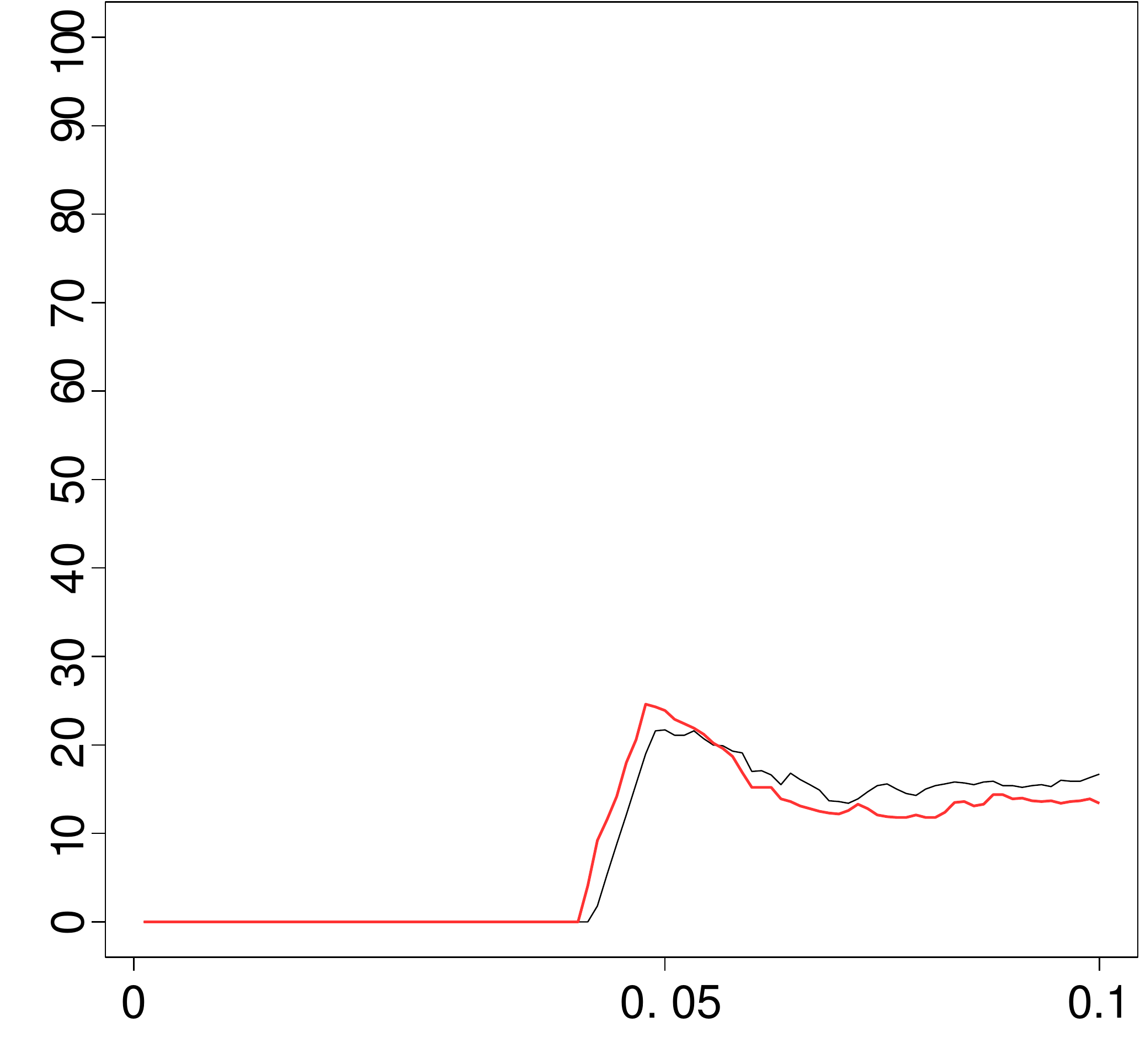}
\includegraphics[width=4cm]{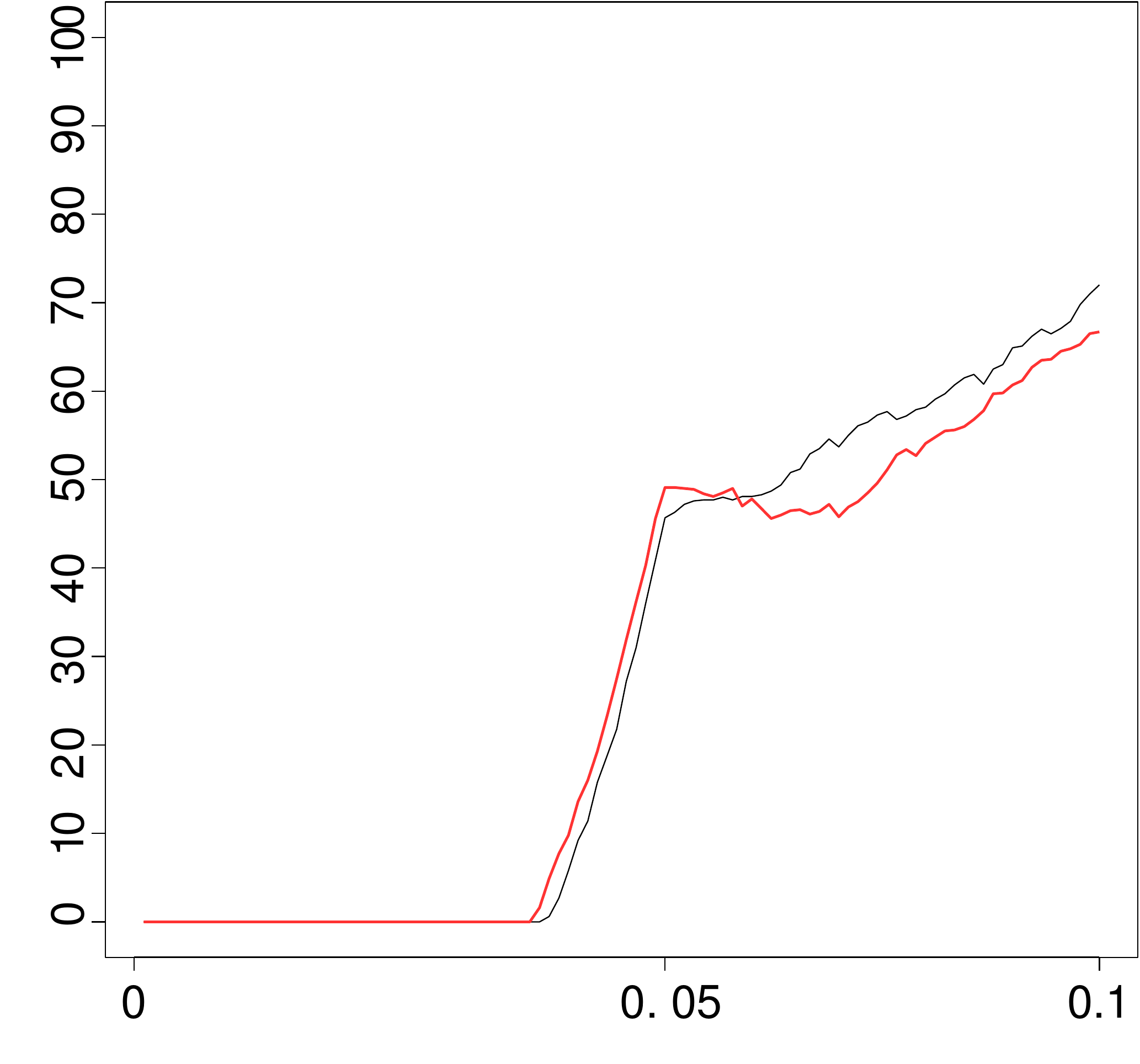}
\includegraphics[width=4cm]{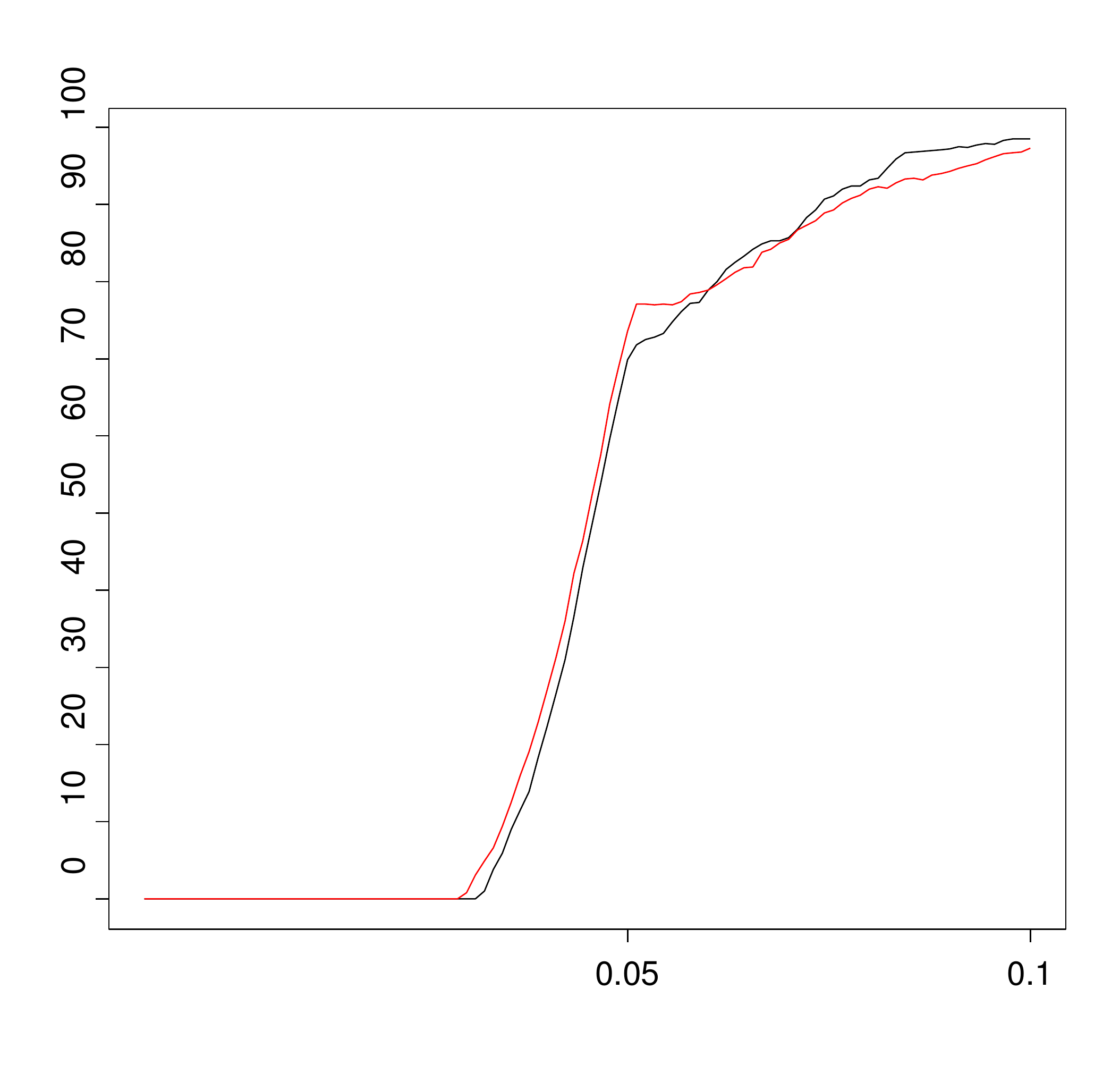}
\caption{The powers of the isotropy test at level 5\% as a function of $r_2$  with respect to $r_\text{cl}$ when using the cylindrical (black) and conical (red)  $K$-function, for the realizations of the PLCPP  with $\sigma = 0.04, 0.02, 0.01, 0.001$ (first four panels from top left to bottom right), Mat\'{e}rn hard-core (third row), and random packing of balls (last row). The last two rows correspond to the factors $c = 0.9, 0.8, 0.7$, from left to right, respectively.
}
\label{fig6:power_both}
\end{figure}

The first four panels of Figure~\ref{fig6:power_both} show the plots of the powers of the isotropy test at a $5\%$ significance level using $m = 1000$ simulations under the PLCPP models with four degrees of linearity as mentioned in Section~\ref{sec:data-plcpp}. The plots indicate that the power of the anisotropy test is slightly higher when using the cylindrical $K$-function than when using the conical one. The shape of the two curves is similar in all plots. In contrast, the last two rows of this figure show that the conical $K$-function is more powerful than the cylindrical one in detecting the anisotropy caused by compression of the regular point patterns when choosing $r_2$ close to the hardcore radius while the cylindrical $K$-function is better for large $r_2$. 

Extra information provided by the plots is that the power of the test obtains its maximum 
where the whole column in the point patterns with columnar structure is captured. As an example to clarify this point, the fourth panel, which is corresponding to a realization of a PLCPP with $\sigma = 0.001$, satisfies our expectation of the diameter of a cylindrical cluster of points to be approximately $4\sigma = 0.004$ (by definition of the PLCPP models). This pattern is followed by the other three values of $\sigma$ as well. In case of the regular point patterns, the maximum is obtained for $r_2$ close to the hardcore radius of $R=0.05$ which corresponds to the findings in \cite{Redenbachetal-09}. 

\begin{figure}[t]
\centering
\includegraphics[width=6cm]{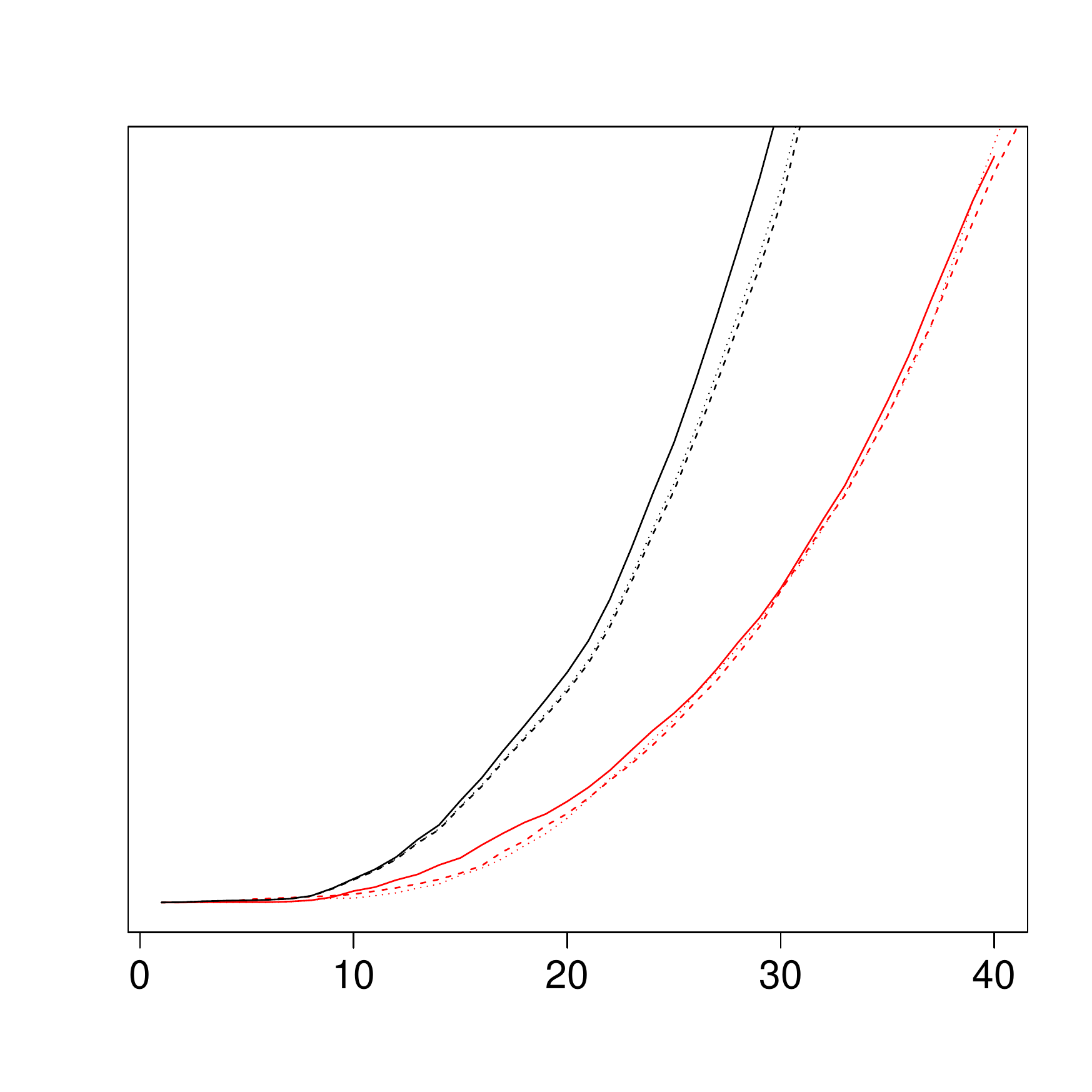}
\includegraphics[width=6cm]{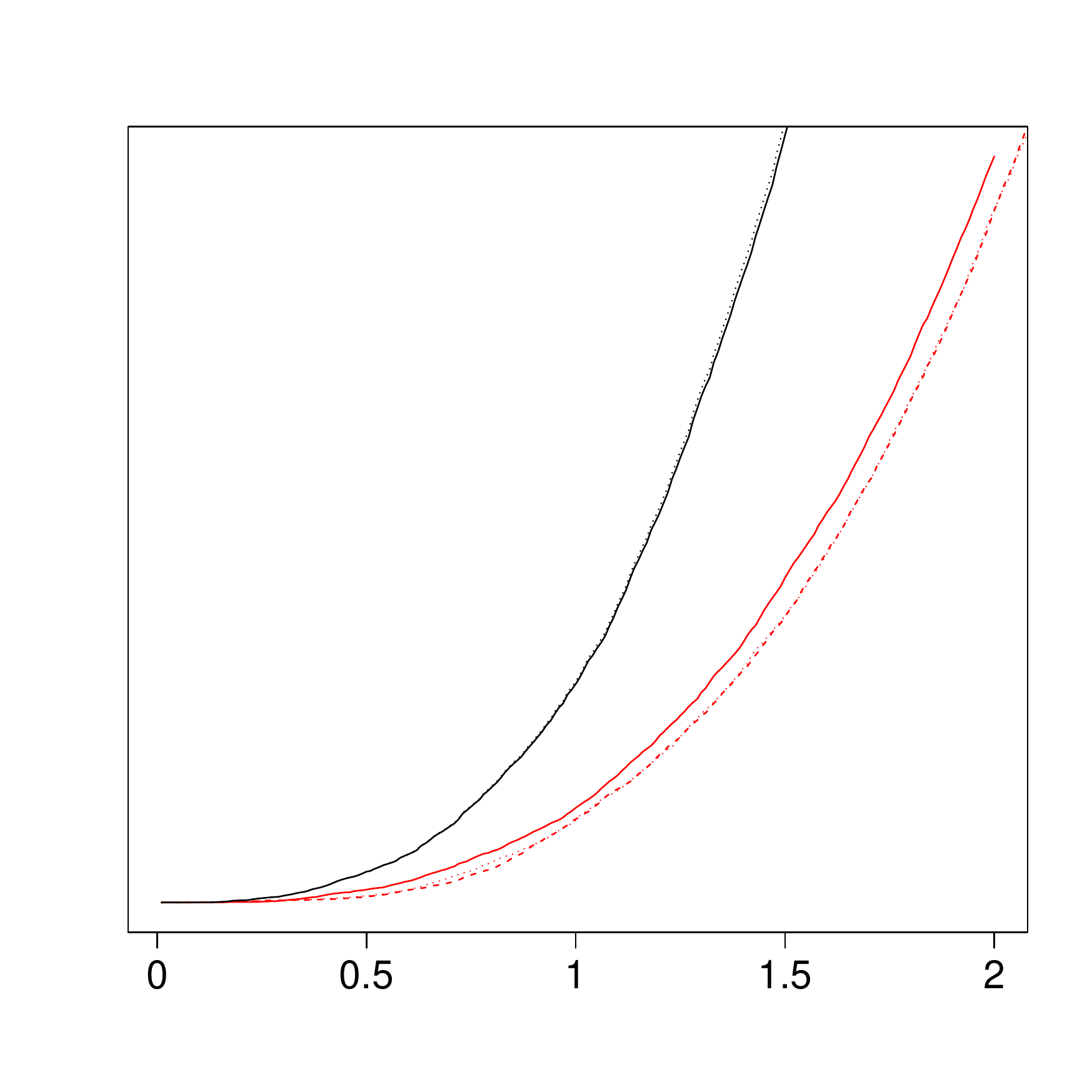}
\caption{Estimates of the cylindrical (black) and the conical (red) $K$-function in the direction of the $z$ (sold), $x$ (dotted), and $y$ (dashed) axis for a sample of the pyramidal cell data (left panel) and the ice data (right panel).}
\label{fig7:KestReal}
\end{figure}

Figure~\ref{fig7:KestReal} shows the estimated $K$-functions for samples of the pyramidal cell and the ice data sets introduced in Sections~\ref{sec:IceData} and \ref{sec:pyramidalData} using the parametrization obtained in equations~\eqref{rh} and \eqref{rr}. As expected based on the power of the isotropy test, the conical $K$-function is more powerful than the cylindrical one in detecting the anisotropy in the ice data. On the other hand, the cylindrical $K$-function is stronger than the conical one in detecting the anisotropy caused by the linear arrangement of the pyramidal cells. Note that we obtained the same behavior when using the rest of samples. 

\section{Discussion}\label{sec:discussion}

In this paper, we have presented a comparison of two directional versions of Ripley's $K$-function using a cone or a cylinder as structuring element. We derived a parametrization to make both functions comparable. Then, both functions were applied to data sets with different sources of anisotropy. The cylindrical $K$-function is generally more powerful than the conical one in case of columnar anisotropy and vice versa in case of compression. In situations where the anisotropy is clearly pronounced, although it can be detected by both functions, the cylindrical $K$-function is clearly more powerful than the conical $K$-function in detecting columnarity.
 
Our application examples show quite different model geometries: points clustered in linear patterns in the minicolumn data and compressed regular point patterns in the ice data. In order to get a comparable testing scenario, we decided to use the nonparametric setting suggested in \cite{Redenbachetal-09}. While this approach is pretty general, it requires replicated data which  are not always available in practice. Nevertheless, an investigation of plots of the directional $K$-functions for different directions may give an indication of existing anisotropies. In cases where a suitable model for the data is available, the test could be replaced by a model based Monte Carlo test. 

The examples given in this paper emphasize the importance of an appropriate choice of the combination of the parameters $(r_\text{cl}, h)$ and $(r_\text{cn}, \theta)$ as well as the integration interval in the test. An unfavorable choice may result in a poor performance of the functions  in detecting the anisotropy of a point pattern. In practical situations, prior information on the construction of the anisotropy, e.g.\ the diameter of the clusters of points in case of the pyramidal cells or the hardcore radius in the regular data, can be used to determine interesting ranges of $r$ values.

Throughout this paper, we assume that the main anisotropy directions are known and fixed. An approach for estimating the main directions in case of the cylindrical $K$-function was discussed in \citet{plcpp-15}. A similar investigation for the ice data has been done in \citet{Rajalaetal-16}. 

\subsection*{Acknowledgments}
This project was supported by the Danish Council for Independent Research | Natural Sciences, grant 12-124675, `Mathematical and Statistical Analysis of Spatial Data', by the Centre for Stochastic Geometry and Advanced Bioimaging, funded by a grant from the Villum Foundation, and by the Center for Mathematical and Computational Modelling (CM)$^\text{2}$ funded by the state of Rhineland-Palatinate, Germany. 

We thank Jens Randel Nyengaard, Karl-Anton Dorph Petersen, and Ali H.\ Rafati at the Center for Stochastic Geometry and Bioimaging (CSGB), Denmark, for collecting the 3D pyramidal cell data, and Johannes Freitag, Alfred-Wegener Institute Bremerhaven, for providing the polar ice data. 

\bibliographystyle{royal}
\bibliography{paper4}
\end{document}